\documentclass[preprintnumbers,eqsecnum,aps,prd,epsf,showpacs,nofootinbib
]{revtex4}
\usepackage{latexsym,amsmath,amssymb}
\usepackage[dvips]{graphicx,color}
\usepackage{dcolumn,bm}    
\usepackage{subfigure}  
\usepackage{color}
\input{colordvi.tex}

\newcommand{\up}{\upsilon }
\begin{document}
\thispagestyle{empty}
\title{Collisions of false vacuum bubbles in cylindrical symmetry}
\author{Yu-ichi Takamizu$^{1}$}
\email{takamizu_at_ccs.tsukuba.ac.jp}
\author{David Chernoff$^{2}$}
\email{chernoff_at_astro.cornell.edu}

\affiliation{$^{1}$ 
Center for Computational Sciences, University of Tsukuba, 
1-1-1 Tennoudai, Ibaragi 305-8577 Japan \\
$^{2}$ 
Department of Astronomy, Cornell University, Ithaca NY 14853\\
}
\date{\today}

\begin{abstract}
  We explore the collision of two cylindrical bubbles in classical
  general relativity with a scalar field stress-energy tensor. Inside
  each bubble the field rests at a local minimum of the potential with
  non-negative energy density. Outside the field rests at zero
  potential, the global minimum.  The calculation resolves the
  connection from the inner de-Sitter region to the asymptotically
  flat Minkowski spacetime. We choose initial conditions such that the
  two bubbles collide and study the full nonlinear evolution by means
  of a two-dimensional numerical simulation of Einstein's
  equations. The collision generates a strongly interacting region
  with spatially varying fields and potentials. These circumstances
  promote dynamical exploration of the potential's landscape. No
  horizon is present and the scalar curvature invariants eventually
  diverge. We speculate that Schwarzschild-like horizons will
  encompass only part of the complicated, interesting regions of
  spacetime in the analogous case of colliding spherical bubbles.
  
\end{abstract}
\maketitle
\section{Introduction}
The string theory landscape \cite{Susskind:2003kw} contains numerous
metastable vacua. The transition paths from one vacuum to another are
of interest in the cosmological history of the universe.  Inspired by
this situation, we investigate the consequences of classical bubble
collisions in Einstein 3+1 gravity for exploring the landscape.

We will focus on scalar field $\phi$ with potential $V(\phi)$ with
action
\begin{equation}
S=\int d^4 x \sqrt{-g} \left[R-\frac{1}{2} \partial_\mu \phi
 \partial^\mu \phi -V(\phi)\right]\,,
\end{equation}
where we set $c=G=1$. We are interested in the situation where
$V(\phi)$ has two or more local minima that permit bubble-like
solutions with true and false vacua. We investigate the outcome of
bubble collisions taking into account self-gravity.

The broad context is the traversal of the potential landscape from
false vacuum to true vacuum and vice-versa. The general statistical
formalism of transition rates \cite{Langer:1969bc} has been coupled
with quantum field theory, including thermal fluctuations, in curved
spacetime \cite{Coleman:1977py,Coleman:1980aw,Hawking:1982my} to
analyze many diverse cosmological scenarios. There are 
situations where the false vacuum to true vacuum transition is of
central interest. The quantum mechanical transition from a de-Sitter
space with positive cosmological constant 
at the termination of inflation in the original inflation models
\cite{Guth:1979bh,Kazanas:1980tx,Sato:1981ds,Guth:1980zm,Guth:1982pn}
is a prominent example. Here, a small region with $V_0>0$ tunnels,
perhaps with thermal assistance, to a lower potential $V_1$ with $V_1
< V_0$. The de-excitation path is of natural interest in an expanding,
cooling universe and the transition rate influences the probability of
bubble collisions and whether the true vacuum percolates in the
expanding background spacetime.

In a different cosmological context, the de-excitation rate governs
how rapidly a complicated theory landscape can be scanned. A primary
question in modern string theory is whether it is possible to locate a
local minima of the multiverse whose potential is consistent with the
small, observed value of the cosmological constant in our Universe
\cite{Polchinski:2006gy}.

Of course, de-excitation transitions from false to true vacua have
corresponding inverse processes. Inside a bubble with small positive
cosmological constant a field patch may fluctuate upward to a local
minima of the potential with larger potential value. While disallowed
in Minkowski space by considerations of energy conservation there are
suggestive general arguments that upward fluctuations will occur in
de-Sitter space \cite{Lee:1987qc}. Fluctuations may also produce
defects during de-Sitter expansion \cite{Basu:1991ig}.

Competing microscopic forward-backward transition rates are important
in the statistical description of nucleation, percolation and
traversal of the landscape.  The possibility of bubble collisions
considerably enriches the variety of transition pathways. Collisions
of true vacuum bubbles may generate short-lived pockets of false
vacuum \cite{Hawking:1982ga} that give birth to black holes.
Energetic bubble collisions can mediate downward transitions and
enhance the rate of exploration of vacua which are close by in field
space \cite{Easther:2009ft}.  Calculations of classical bubble
collisions have been carried out for various model potentials
\cite{Giblin:2010bd,Hwang:2012pj,Hwang:2014cqa}. The observational
signatures of bubble collisions are also being explored
\cite{Aguirre:2007an,Aguirre:2009ug,Wainwright:2013lea,Wainwright:2014pta,Johnson:2015gma,Kleban:2011pg}.

In this paper we will concentrate on collisions of bubbles of false
vacua. We will assume that one bubble collides with another bubble
before either is hidden by an apparent horizon and before either expands
beyond the cosmological horizon. We consider the simplest scalar field
potential with one global minimum with $V=0$, two distinct local
minima with $V>0$ and intervening barriers. This choice allows the
study of many types of bubble collisions.

\section{Metric for spherical and cylindrical bubbles}
In 3+1 dimensional spacetime with spherical symmetry the metric with
isotropic coordinates may be written $ds^2 = -\alpha^2 dt^2 + a^2
(dr^2 + r^2 d\Omega^2$) where $r$ is a radial coordinate,
$\alpha=\alpha(t,r)$, $a=a(t,r)$ and $d\Omega^2$ is the
two-dimensional metric on the unit sphere (see eqn. 16 in
\cite{Parry:2012ku}). Spherical symmetry is suitable for an isolated
bubble and the evolution can be handled numerically since the relevant
dynamics is 1+1 dimensional.  Two spherically symmetric bubbles can
collide with axial symmetry about the line of separation of their
centers but the numerical solution is 2+1 dimensional and more
challenging. More general collisions lead to full 3+1 dimensional
problems.  Instead, we will consider bubbles of a cylindrical form
that vary in space perpendicular to the axis of symmetry but not along
the axis.  The dynamics of a single bubble is 1+1 dimensional and the
collision of two bubbles with aligned axes is 2+1 dimensional. We
adopt this simplified geometry as a first step in the study the
interactions of bubbles.

The metric for a single cylindrically symmetric bubble has the form
\begin{eqnarray}
  ds^2 = e^{2A}\left(-dt^2+dr^2+r^2 \tilde{\beta}^2 
d\theta^2\right) +e^{2(B-A) }dz^2, 
\label{eq-metric-cylinder-system}
\end{eqnarray}
where the three metric functions $A$, $B$ and $\tilde{\beta}$ depend
upon $t$ and $r$ (the two dimensional cylindrical radius).  This form
is equivalent to the general expression given by Thorne (eqn. 2 in
\cite{Thorne65}). We use $\theta$ as the polar angle since we will
retain the symbol $\phi$ for the scalar field potential. All functions
depend upon $t$ and $r$ but not $\theta$. This metric has axial
symmetry about the z-axis and translational symmetry along the z-axis
and we refer to it as a ``cylinder system''.

The deficit angle for the cylinder system is $\delta$, where
$\tilde{\beta}=(2\pi-\delta)/2\pi$. The simplest example of the
deficit angle occurs in an ideal one-dimensional cosmic string with
tension $\mu$. Solving Einstein's equations gives $\delta = 8 \pi \mu$
constant everywhere and ill-defined at $r=0$.  In general, the deficit
angle is related to the mass-energy distribution within the cylinder
system and varies with $r$. We will work in the limit that $\delta$ is
initially small and remains small.  The dynamical equations for
$\tilde{\beta}$ are of the form
\begin{eqnarray}
  \tilde{\beta}_{tt} =
 \tilde{\beta}\times {\cal C} -B_r \tilde{\beta}_r+ 
B_t \tilde{\beta}_t, 
\label{eq-beta-dynamics}
\end{eqnarray}
where ${\cal C}$ is schematic for the left-hand side of the constraint
eq. \ref{const1} below. When the constraint equation is satisfied
${\cal C}=0$. So, if $\tilde{\beta}_r=\tilde{\beta}_t=0$ initially and
if the constraint equation is exactly imposed then $\tilde{\beta}=1$
and $\delta=0$. Henceforth, we drop the spacetime dependence for
$\tilde{\beta}$, fixing $\tilde{\beta}=1$.

Now the form of the metric for the cylinder system becomes
\begin{eqnarray}
  ds^2 &=& e^{2A}\left(-dt^2+dx^2+dy^2 \right) +e^{2(B-A) }dz^2, \\
&& A = A(r,t), ~~B = B(r,t), 
\label{eq-metric}
\end{eqnarray}
where $dr^2+r^2d\theta^2=dx^2+dy^2$. For a single bubble the maximally
symmetric metric form has $B=2A$. Numerical calculations show that an
initial state of maximal symmetry evolves in such a way that
the symmetry is broken but that the deviation is small.  Later,
we will find that this remains true even for two interacting bubbles.

Up to now we have considered a single cylinder system with axial
symmetry.  For two bubbles the axial symmetry is broken.  Assume the
bubble deficit angle is small, the metric approaches Minkowski and the
scalar field is uniform far from the cylinder.  For two distant,
parallel cylinders of this sort at rest with respect to each other we
promote the two separate $A$ functions (one for each cylinder) to a
single function $A$ for all spacetime.  Likewise for $B$.  The
promoted $A$ and $B$ will depend upon $r$, $\theta$ and $t$, or
equivalently, $x$, $y$ and $t$. For close, interacting cylinders
(either because the outer geometry is not Minkowski or because the
scalar field varies) our working ansatz is to replace the occurrences
of $A(r,t)$ and $B(r,t)$ in the cylinder system with the promoted
$A(x,y,t)$ and $B(x,y,t)$, respectively, while holding the form of the
metric fixed. We continue to assume $\delta = 0$ in the interacting
system.  This ansatz yields three evolution equations and five
constraint equations that extend the original three evolution
equations and three constraints of the cylinder system.  The
hierarchical relationship of extended to original systems is
straightforward: when $A(x,y,t) \to A(r,t)$, $B(x,y,t) \to B(r,t)$ and
$\phi(x,y,t) \to \phi(r,t)$ the evolution equations for
$A_{tt}(x,y,t)$, $B_{tt}(x,y,t)$ and $\phi_{tt}(x,y,t)$ reduce to the
equivalent expressions for $A_{tt}(r,t)$, $B_{tt}(r,t)$ and
$\phi_{tt}(r,t)$, respectively. Likewise, the five constraints reduce
to the three independent constraints of the cylinder system.

Typically, the initial conditions we choose have maximal symmetry but
the system can/will evolve away from maximal symmetry, i.e. $B=2A$
is not imposed.  We generalize the scalar field $\phi(r,t)$ to $\phi(x,y,t)$.

\section{Potential}

\begin{figure}[htbp]
\begin{center}
\includegraphics[width=6cm]{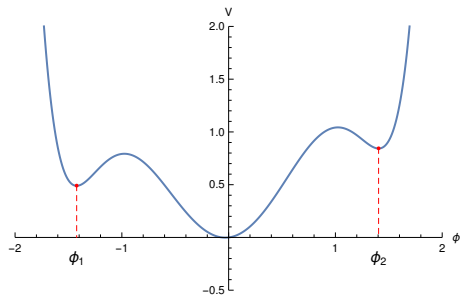} 
\end{center}
\caption{The scalar field potential $V(\phi)$, shown for
  $b_0=1$, possesses 2 local minima with positive cosmological
  constant and 1 global minimum with zero cosmological constant. The
  transition of field from the global minimum (Minkowski spacetime) to
  one of the local minima (de-Sitter spacetime) by way of the local
  maximum forms the bubble wall.}
\label{fig-BCV}
\end{figure}

We will consider a potential that allows for two bubble vacua at
$\phi=\phi_1$ and $\phi_2$ with $V(\phi_1)>0$ and $V(\phi_2)>0$. These
will collide in a background
region with $\phi=\phi_0$ with $V(\phi_0)=0$. The specific form is
\begin{equation}
V(\phi)=b_0\left({\frac{1}{12}}(\phi^4-1)^2-(\phi^2-1)^2+{\frac{1}{8}}(\phi-1)+c_0\right)\,.
\end{equation}
The global minimum of the potential occurs at $\phi=\phi_0=-0.031$ and
the constant $c_0$ sets $V(\phi_0)=0$. The shape of the potential is
shown in Fig. \ref{fig-BCV}. The two local minima $\phi_1$ and
$\phi_2$ have the following properties:
\begin{equation}
V(\phi_1)<V(\phi_2),~~ \phi_1=-1.42,~~\phi_2=1.40\,.
\end{equation}
The cosmological constant for each bubble is
$\Lambda_{1,2}=V(\phi_{1,2})$.  The multiplicative factor $b_0$ sets
the scale of the potential.

In the limit of uniform, time-independent $\phi$ the conformal Hubble
constants (the physical time is $\tau=\int e^{A} dt$; the conformal
time is $t$) are
\begin{equation}
H_{1,2}\equiv \partial_t A_{1,2}=e^{A_{1,2}}\sqrt{\frac{V(\phi_{1,2})}{3}}\,
\label{Hubble-eq}
\end{equation}
for bubbles $1$ and $2$.  The rate of expansion is proportional to
$\sqrt{b_0}$.

The Einstein equations and the scalar field equation of motion yield
the dynamical equations
\begin{align}
4A_{tt}(t,x,y)&=-2A_t^2+4 (A_{xx}+A_{yy})+2(A_x^2+A_y^2)-(\phi_t^2-\phi_x^2-\phi_y^2)+2e^{2A}V(\phi)\,,\label{dyn1}\\
2B_{tt}(t,x,y)&=B_{xx}+B_{yy}-(A_x^2+A_y^2)+2(A_x B_x+A_y B_y)+
(A_x-B_x)^2 +(A_y- B_y)^2\nonumber\\
& -2(A_t-B_t)^2-\phi_t^2+2e^{2A}V(\phi)\,,\label{dyn12}\\
\phi_{tt}(t,x,y)&=-B_t\phi_t+\phi_{xx}+\phi_{yy}+B_x\phi_x +B_y\phi_y-e^{2A}\partial_\phi V(\phi)\,,\label{dyn2}
\end{align}
plus constraint equations
\begin{align}
&B_{xx}-B_{yy}+A_x^2-A_y^2+2A_y B_y-2A_x B_x+
(A_x-B_x)^2-(A_y-B_y)^2+\phi_x^2-\phi_y^2=0\,,\label{const1}\\
&2A_t(2B_t-A_t)=2(B_{xx}+B_{yy})+2(A_x-B_x)^2+2(A_y-B_y)^2+\phi_t^2+
\phi_x^2+\phi_y^2 +2e^{2A}V(\phi)\,,\label{const-H}\\
&B_x(2A_t-B_t)+2A_x(B_t-A_t)-B_{xt}=\phi_x\phi_t\,,~~B_y(2A_t-B_t)+2A_y(B_t-A_t)-B_{yt}=\phi_y\phi_t\,,\label{const2}\\
&B_y(2A_x-B_x)+2A_y(B_x-A_x)-B_{xy}=\phi_x\phi_y\,,\label{const3}
\end{align}
where $A_t$, $A_x$ and $A_y$ denotes partial derivatives of $A$ with respect to
$t$, $x$ and $y$, respectively.  The constraint equations are
solved for the initial conditions and, once satisfied, are
automatically preserved by evolution.  We refer to equation (\ref{const-H})
as the FLRW equation for the whole spacetime including two bubbles.

\newpage

\section{Non-interacting bubble solutions}
First we will consider two thin-walled bubbles with stationary centers
that lie far apart. We work in a two-dimensional box with coordinates
$x=[-20,20]$ and $y=[-20,20]$ and take the centers of two bubbles to
be $(x_1,y_1)=(-7,0)$ and $(x_2,y_2)=(7,0)$. The initial bubble radii
are $R_{1,2}=4$ where the bubble rim in $(x,y)$ space satisfies
\begin{equation}
R_{1,2}=\sqrt{(x-x_{1,2})^2+y^2}\,.
\end{equation}

We will specify the initial metric and scalar field configuration by
using the above equations. For the metric, we fix $B=2A$ at the
initial time (maximal symmetric case).  Eqns. (\ref{dyn1}) and
(\ref{dyn12}) yield spatial differential equations without any time
derivative terms.  Holding $\phi_t=\phi_{tt}=0$ and using (\ref{dyn2})
gives
\begin{align}
2(A_{xx}+A_{yy})-2(A_x^2+A_y^2)+\phi_x^2+\phi_y^2=0\,,\label{basic-static0}\\
\phi_{xx}+\phi_{yy}+2A_x\phi_x +2A_y\phi_x-e^{2A}\partial_\phi V(\phi)=0\,.
\label{basic-static}
\end{align}
An isolated bubble has zero field gradient at the bubble center and
the field approaches zero at infinity. We approximate by setting
$\phi=\phi_i$ and $A=A_i$ within bubble $i$ ($i=1$ and $2$) and
$\phi=A=0$ on the computational box boundaries.  Numerically, we
choose $A_1=0.25$ and $A_2=0.3$. Once $A>{\cal O}(1)$ large changes in
the metric will occur on account of the appearance of the factor
$e^{2A}$.

\begin{figure}[htbp]
\begin{center}
\begin{tabular}{ll}
\includegraphics[width=8.5cm]{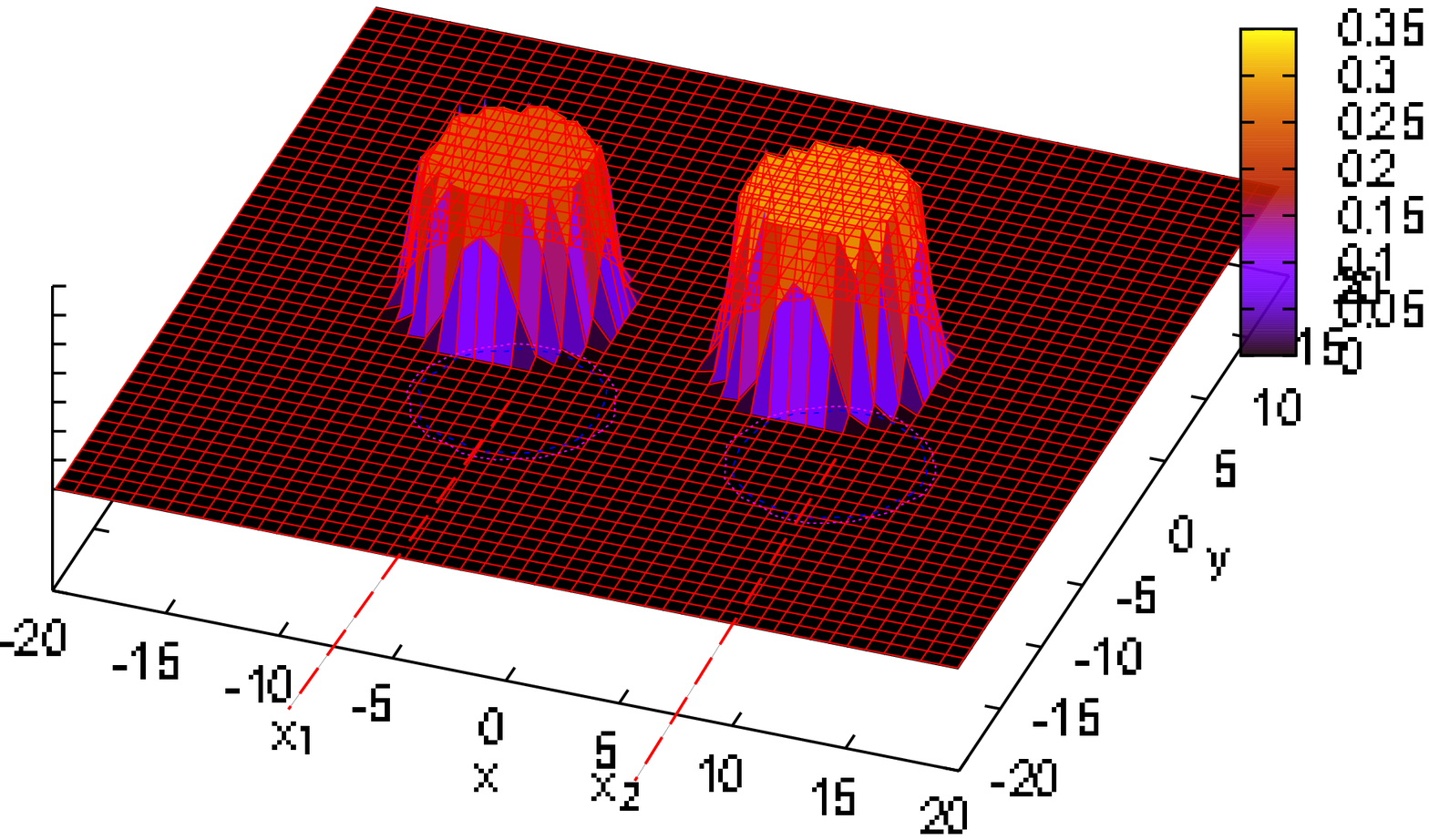} &
\hspace{0.5cm}
\includegraphics[width=8.5cm]{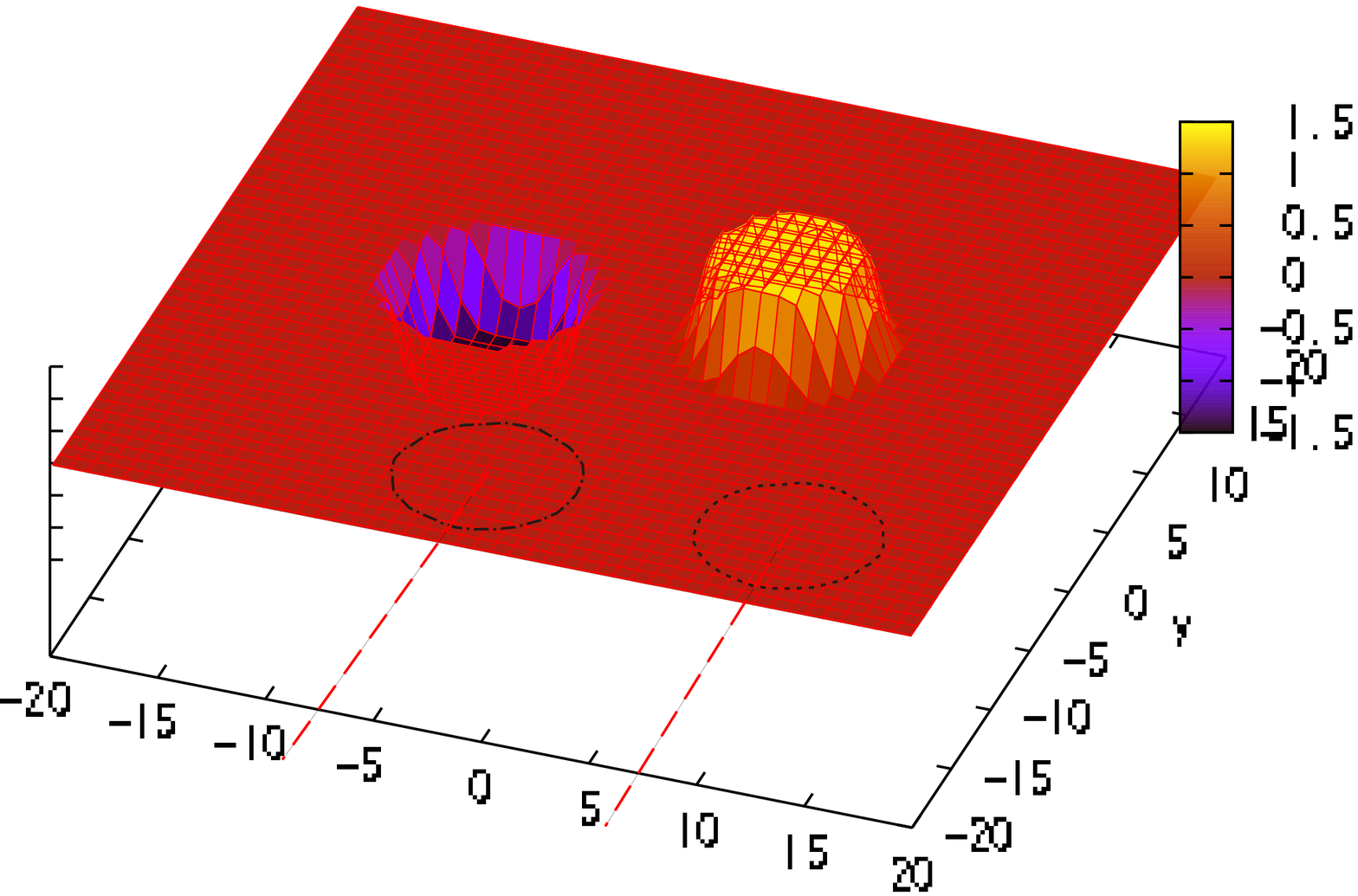}
\end{tabular}
\end{center}
\caption{Snapshot of metric coefficient $A(x,y,t)$ (left) and field
  $\phi(x,y,t)$ (right) in the $x$-$y$ plane perpendicular to the axis
  of symmetry at initial time $t=0$. The surface plots of $A$ and
  $\phi$ are also color coded by value.  The red dashed lines below
  the surface mark the $x$ coordinate of the center of the bubble.
  The circular lines below the surface trace the bubble rims.
}
\label{fig-nonbst3}
\end{figure}

The numerical solution is based on approximating the functions
$A(t,x,y)$, $B(t,x,y)$ and $\phi(t,x,y)$ at a given time by function
values on a grid with points separated by $\Delta x=\Delta y=0.01$ of
total size $4001 \times 4001$.  At $t=0$ the nonlinear constraint
equations are converted to finite difference form and solved using
Newton-Raphson techniques while holding the region within the rim and
along the computational boundaries fixed. Satisfying equations
(\ref{basic-static0}) and (\ref{basic-static}) in cylinder symmetry in
the $(x,y)$ plane automatically satisfies the individual equations in
(\ref{const1}) and (\ref{const3}). The initial spatial solutions for
metric and scalar field of two bubbles (different vacua) are
illustrated in Fig. \ref{fig-nonbst3}.

Once we have found the initial spatial solutions $A(t=0,x,y)$ and
$\phi(t=0,x,y)$, we infer the initial time derivatives $A_t(t=0,x,y)$
and $\phi_t(t=0,x,y)$ by solving the other constraint equations
(\ref{const-H}) and (\ref{const2}).  There are effectively two
equations for two variables $A_t$ and $\phi_t$ since the second
equation of (\ref{const2}) is same as the first one due to symmetry of
the isolated bubble. At the box edge, far outside the bubble, we
impose $A_t=\phi_t=0$ (flat Minkowski space, stationary field at the
global minimum) and inside the bubble we impose $\phi_t=0$ (stationary
field with the vacuum expectation value). We infer the Hubble
expansion of the bubble (\ref{Hubble-eq}) from the FLRW equation
(\ref{const-H}). The transition near the bubble wall has
non-trivial initial time derivatives in the metric and field that
follow from the above choices. We do not impose
these quantities.

\begin{figure}[htbp]
\begin{center}
\begin{tabular}{ll}
\includegraphics[width=6cm]{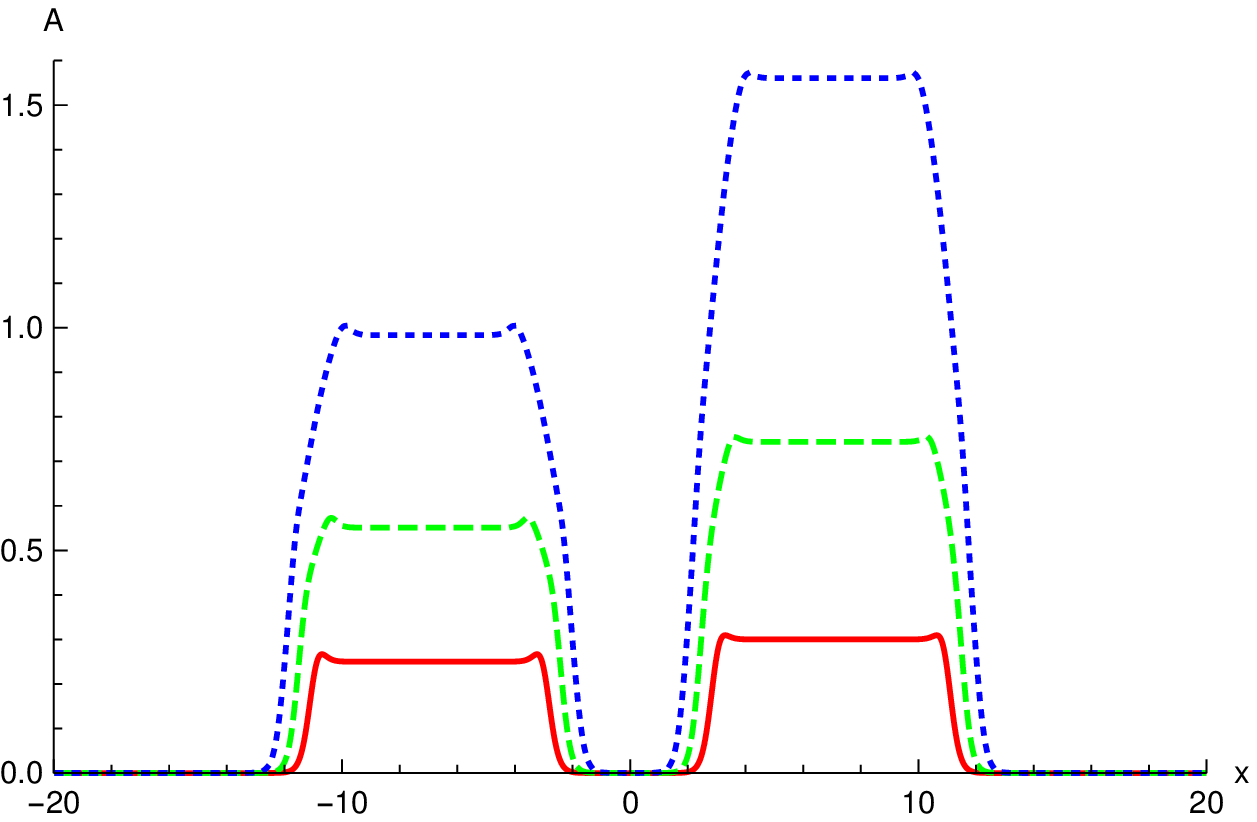} &  
\hspace{0.5cm}
\includegraphics[width=6cm]{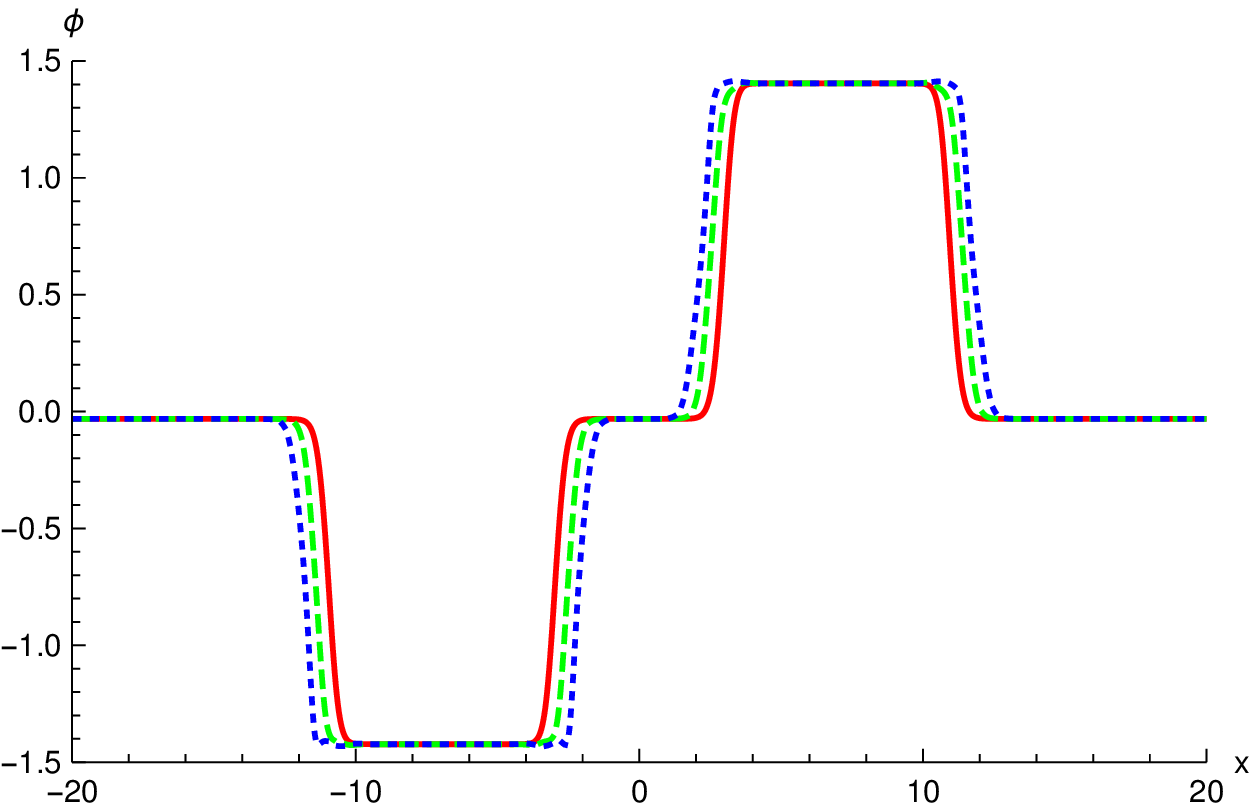}
\end{tabular}
\end{center}
\caption{The figure shows the evolution of the metric function $A$
  (left) and scalar field $\phi$ (right) for spacetime with two
  initially isolated bubbles. The bubble at $x<0$ has field
  $\phi=\phi_1$, the one at $x>0$ has $\phi=\phi_2$. The bubbles are
  surrounded by Minkowski spacetime with $\phi=\phi_0$. The bubble
  rims are symmetric in $y$ about $y=0$.  The left panel gives three
  successive one-dimensional cuts of the metric function, $A(t,x,y=0)$
  for $t=0$ (red), $0.5$ (green) and $1$ (blue). The bubble interiors experience
  de-Sitter-like expansion.  Since $V(\phi_2) > V(\phi_1)$ the growth
  of $A$ on the right is larger than on the left. The background
  spacetime has constant $A$. The right panel shows $\phi(t,x,y=0)$
  along the same one-dimensional cut and at the same set of times. The
  field transition from one minimum to another is numerically
  well-resolved in the calculations. Over the short time interval plotted,
  each wall expands outward at approximately constant, relativistic
  speed. The two bubbles will eventually interact. These calculations
  use $b_0=1$.}
\label{fig-nonbst1}
\end{figure}

With the initial values ($A$, $B=2A$ and $\phi$) and time derivatives
($A_t$ and $\phi_t$) we solve the dynamical equations (\ref{dyn1}),
(\ref{dyn12}) and (\ref{dyn2}) for time evolution (see
Fig. \ref{fig-nonbst1}).  The internal metric grows in a manner
consistent with the choice of the local cosmological constant. Within
the
bubbles, $A_{1,2}(t) \simeq H_{1,2} t+A_{1,2}|_{t=0}$ at early times
$t<{\cal O}(1)$.  Later, $A_{1,2}$ grows more rapidly.  We find that
the metric coefficient $B$ approximately tracks $A$ in the sense $B
\sim 2A$ (not illustrated in the figure).  More precisely, $B < 2A$ at
later time and the difference $|B-2A| < 10^{-2}$ is small.  Maximal
symmetry breaks and manifests as the shrinkage of the ratio of the
length in $z$-direction to that in the $x$-$y$ directions. The ratio
is $\propto e^{2(B-2A)}$.

Fig. \ref{fig-nonbst1} also shows that each bubble expands on account
of the choice of initial conditions. The plots show an advancing time
sequence of profiles of the metric function $A$ and the scalar field
$\phi$ before the individual bubbles begin to interact.  The right
hand plot shows that the rim of the bubble expands with coordinate
speed ${\dot R} \sim 0.8-0.9$ (where the speed of light is 1).  The
field transition from inside to outside the bubble is well-resolved by
the numerical grid. Its detailed shape is determined by the form of
the scalar potential that connects the two minima.

\begin{figure}
\begin{center}
\begin{tabular}{ll}
\includegraphics[width=6cm]{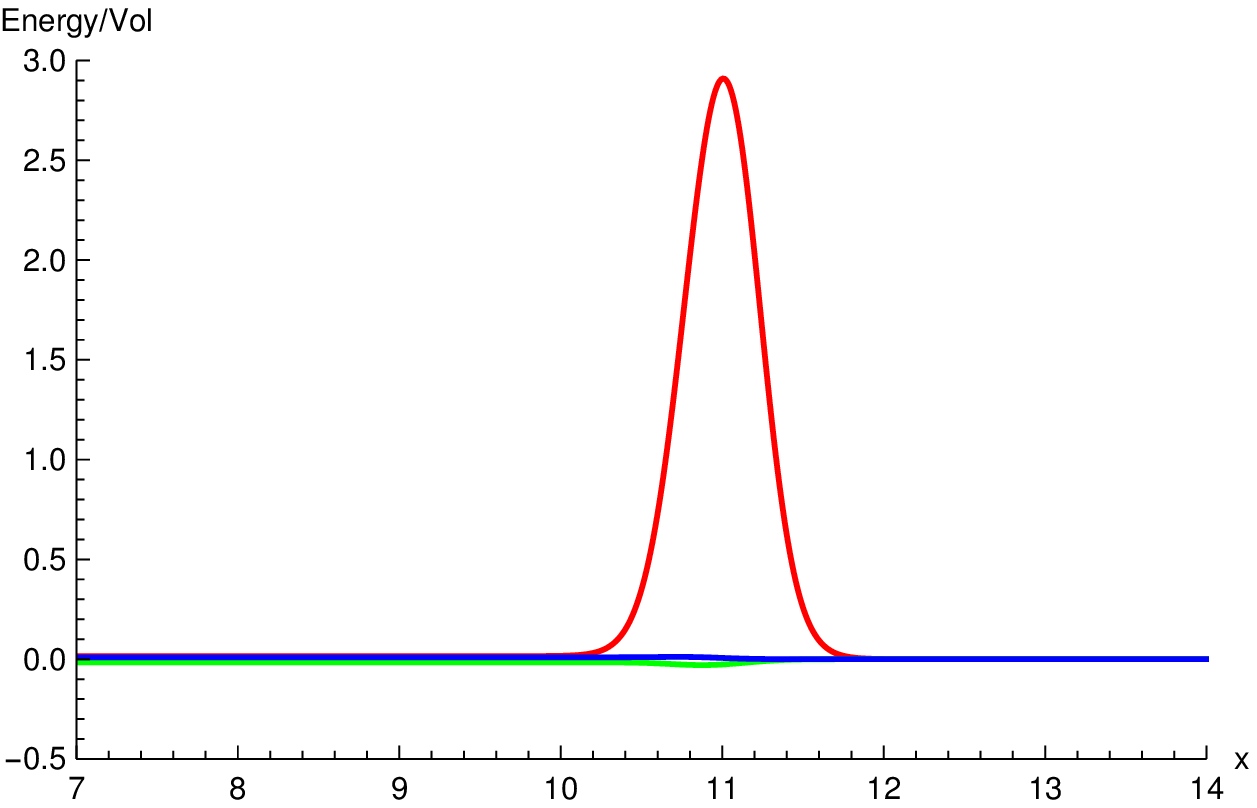} &
\hspace{0.5cm}
\includegraphics[width=6cm]{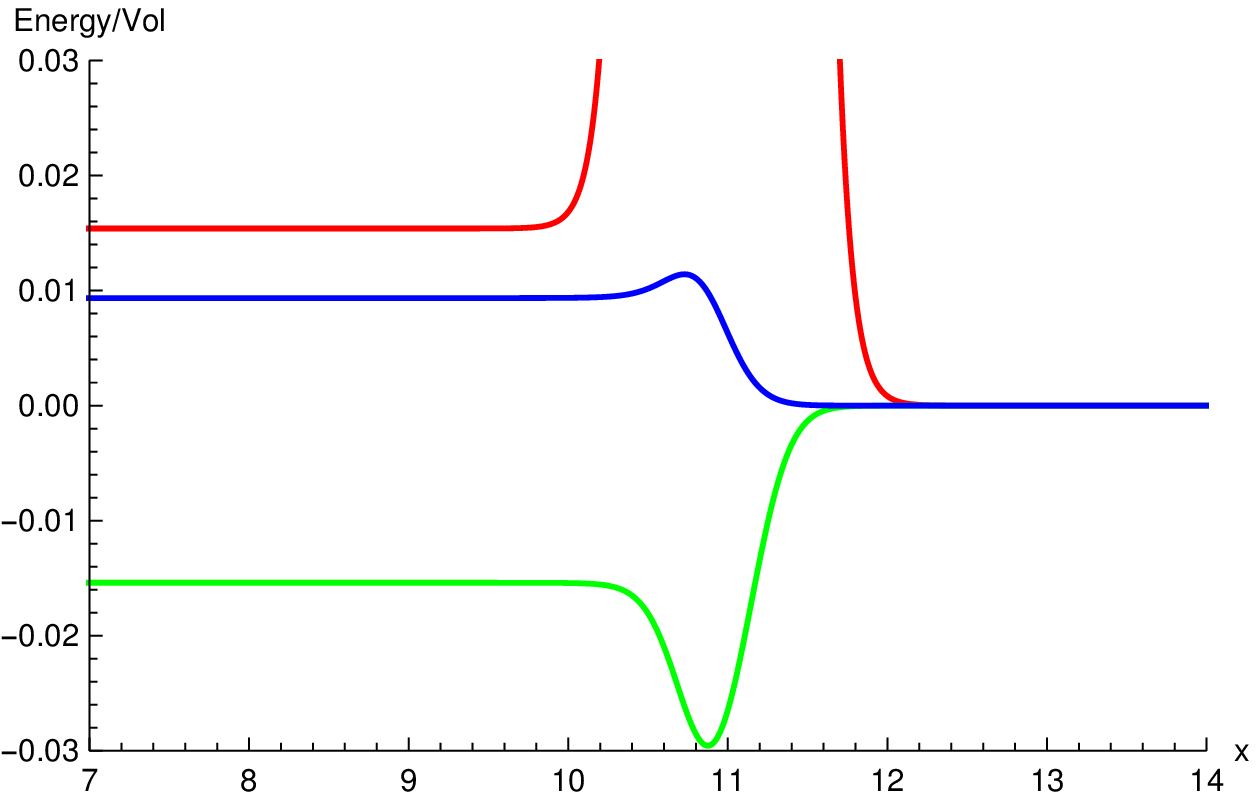}
\end{tabular}
\end{center}
\caption{The figure shows the energy density $\rho$ (red), the
  pressure $P$ (green) and the metric contribution $A_t^2$ (blue) for
  the bubble near the beginning of the simulation. The left hand
  figure shows the full range of variation. The right hand figure
  shows an expanded view of the smaller contributions near the shell.
  \label{forces}}
\end{figure}

Three physical effects control the evolution: the energy density
inside the bubble (that of the false vacuum) and the surface tension
of the bubble (as the scalar field surmounts the barrier between false
and true vacua) are attractive. These pull the wall toward the center
of symmetry.  Within the bubble the negative pressure dominates the
FLRW equation and drives the interior's exponential
expansion. Finally, the jump in pressure at the wall (from inside to
out) creates an additional inward directed force on the
wall. Fig. \ref{forces} shows the local density $\rho = (1/2)
(\phi_t^2 + \phi_x^2 + \phi_y^2) + e^{2A} V$, pressure $P = (1/2)
(\phi_t^2 - \phi_x^2 - \phi_y^2) - e^{2A} V$ and metric contribution
$A_t(2B_t - A_t)/3 \simeq A_t^2$ as they appear in the FLRW equation.
The left hand plot shows the dominance of the energy density within
the transition region of the shell. By comparison, the interior
pressure is small but dominant within the interior. The right hand
plot shows that $\rho$, $P$ and $A_t^2$ are comparable within the
false vacuum bubble.
Based on the scale of the contributions we expect the shell to collapse
even while the interior inflates.

Many investigations have characterized the general relativistic
dynamics of models possessing a thin shell with surface tension that
separates a spherical false vacuum interior from an asymptotically
flat exterior.
\cite{Berezin:1982ur,Maeda:1985ye,Sato:1986xxxx,Blau:1986cw,Berezin:1987bc,Aurilia:1989sb}. Suzuki
et al \cite{Suzuki:1991yk} provide a solution for a thin
cylindrical bubble with false vacuum interior and flat exterior. Their
coordinates differ from ours. In their description the rim motion is
equivalent to particle motion in a potential. They found the simple
result that the bubble wall which begins at small radius cannot reach
infinity but must fall back to the origin. 

To compare our solution quantitatively with theirs we transformed our
initial coordinates for the bubble rim with respect to the bubble
center ($\Delta r=6$ and coordinate velocity, ${\dot r}=dr/dt=1$) to
their system of coordinates, solved for the rim motion and transformed
the result back to our own system.
\footnote{Suzuki's coordinates are $\rho$ and $T$. Here, $\rho$ is
  measured about the line of symmetry. Ours are $\Delta r$ and $t$
  where $\Delta r$ is $r$ measured about the bubble center.  The
  relationship is $\rho = \int e^A dr$ and $T = \int e^A dt$ where
  $A=A(t)$ has been taken to be only a function of time inside the
  bubble. We inferred the surface tension for our potential by direct
  numerical integration of the false to true vacuum transition from
  interior to exterior $ \sigma = \int \rho dr$ where $\rho =
  \frac{1}{2}{\dot \phi}^2 + e^{2A} V(\phi)$. This gives $G \sigma =
  1.32$. The corresponding terms for pressure and metric yield $-2.8
  \times 10^{-2}$ and $6.4 \times 10^{-3}$ respectively.  We solved
  for the motion $\rho(T)$ using their eqns. 3.7a-c and then
  transformed the results back to $\Delta r(t)$ in our own system.}
The plot shows the Suzuki solution for the bubble rim and the rim
position inferred from the maximum total energy density in our
coordinate system. In both systems the false vacuum bubble collapses.
\begin{figure}[htbp]
\begin{center}
\includegraphics[width=12cm]{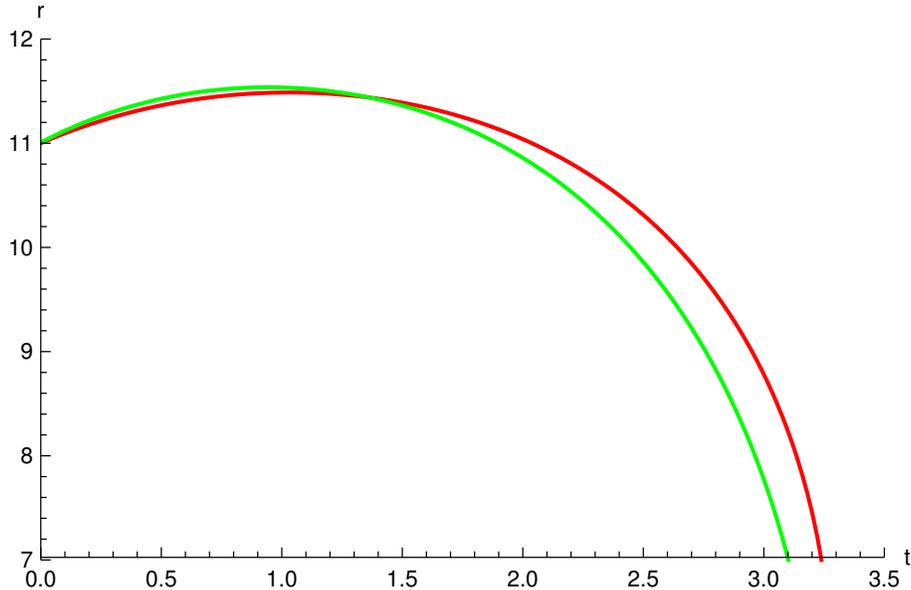}
\end{center}
\caption{A comparison of the shell position based on Suzuki's thin
  shell solution (red) and a numerically calculated (green) bubble
  like those on the right side of fig. \ref{fig-nonbst1}. The bubble
  center is at $x=7$ and the initial radius is $4$. The rim position
  in the numerical simulation is based on the maximum total energy
  density.
\label{isolated-bubble}}
\end{figure}

In our coordinates the positive, nearly constant potential within the
bubble implies fixed Hubble value within. Physical separation grows in
an exponential fashion.  As discussed by
\cite{Sato:1981gv,Blau:1986cw,Berezin:1987bc} for spherical bubbles,
an observer in the false vacuum expects to see inflation whereas one
near the transition between false and true vacuum (assumed thin)
expects to see that inward directed forces govern the wall motion.  In
our case, the thickness of the wall becomes important. A gradient in
the Hubble constant (rate of inflation) inevitably appears. In our
computational coordinates, the inner part of the wall traces the thin
shell's motion. It surrounds the nearly flat, false vacuum interior
and collapses in the manner described by the Suzuki solution. The
interior region is inflating at the maximum rate. The outer parts of
the wall move outward with coordinate velocities close to 1 and
inflate at smaller rates. The transition shears the wall into a new,
bridge-like region with varying Hubble expansion.
This phenomena also implies that the two neighboring bubbles in
fig. \ref{fig-nonbst1} will begin to interact even though the rims
would not meet in the thin wall treatment.

\begin{figure}[htbp]
\begin{center}
\includegraphics[width=6cm]{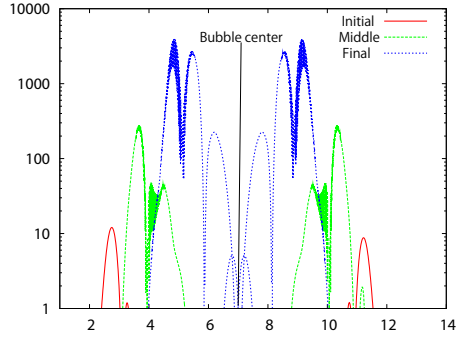}
\end{center}
\caption{The Kretschmann scalar as a function of position at three
  times $t=0$, $2.4$ and $2.7$ during the simulation, labeled initial,
  middle and final respectively in the plot. The diverging curvature tracks
  the rim position of the Suzuki solution.}
\label{fig-RRx}
\end{figure}

\begin{figure}[htbp]
\begin{center}
\begin{tabular}{ll}
\includegraphics[width=7cm]{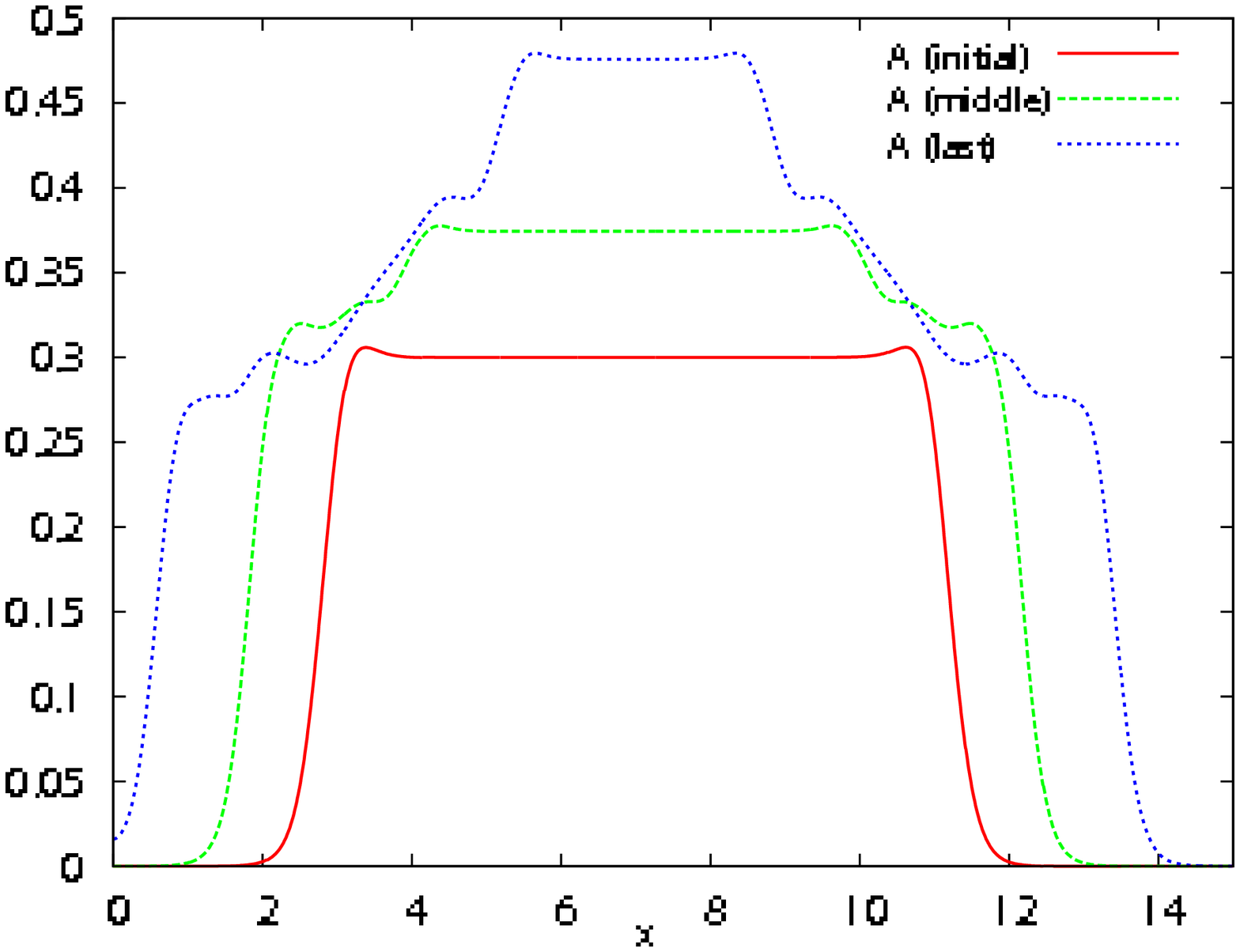} &
\hspace{0.5cm}
\includegraphics[width=8.5cm]{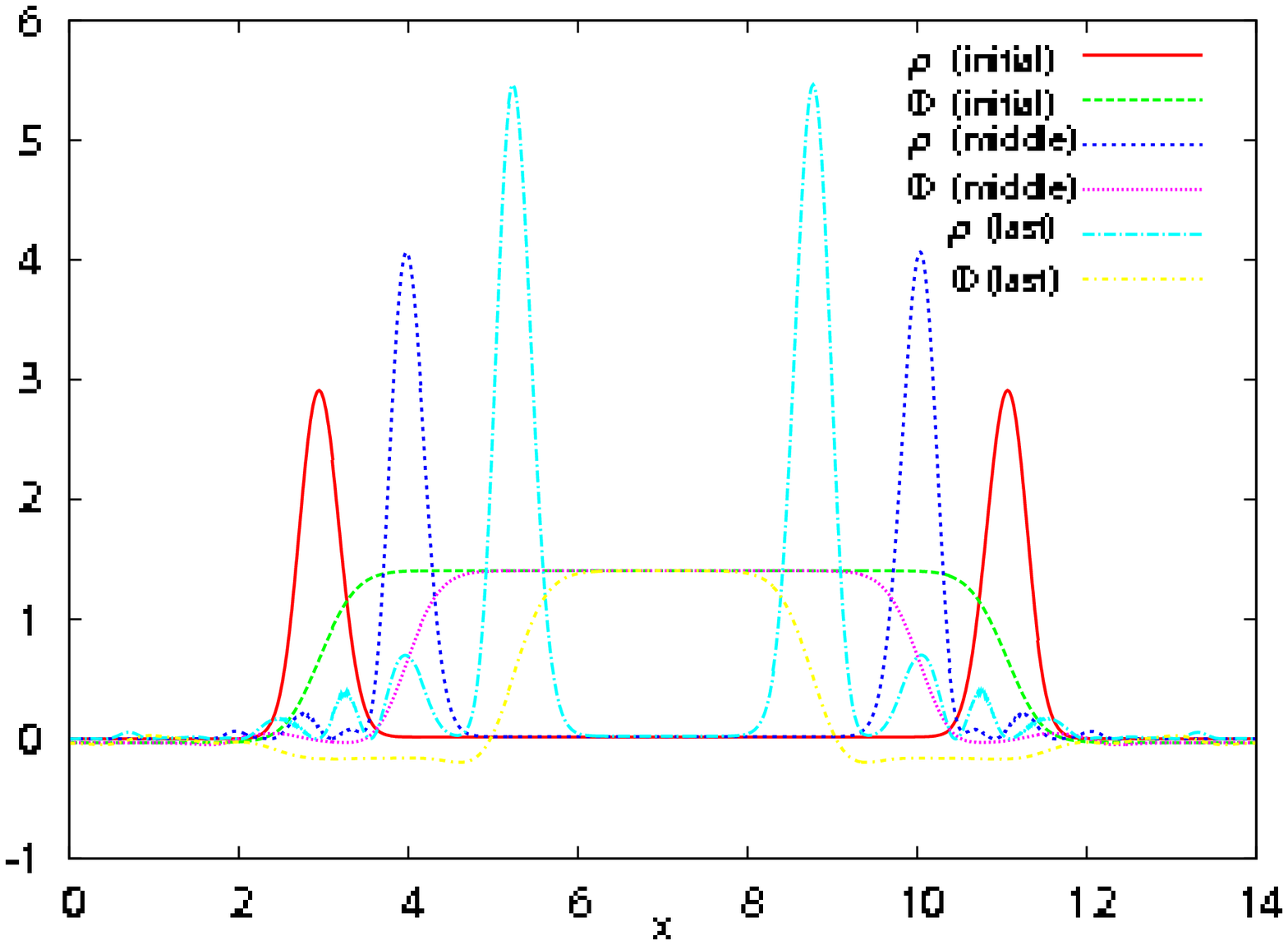} 
\end{tabular}
\end{center}
\caption{The time evolution of the metric $A$, scalar field $\phi$ and
  energy density $\rho$ for one contracting single bubble. The
  initial, middle and final labels refer to the same times as in
  Fig. \ref{fig-RRx}. The left panel shows that the metric increases
  inside while the wall region which separates the inside from outside
  diffuses in both directions. The right panel shows the peak of the
  energy density and the point of the scalar field transition move
  inward with time. The maximum energy density tracks the thin
  rim location in the Suzuki model.}
\label{fig-RRap}
\end{figure}

Fig. \ref{fig-RRx} plots the Kretschmann scalar curvature
($K=R^{abcd}R_{abcd}$) at three successive moments of time for the
bubble on the right. The metric, scalar field and energy density are
shown in Fig. \ref{fig-RRap}.
In a de-Sitter space with Hubble constant $H$ the Kretschmann scalar
$K=24 H^4$.  This is equivalent to $K = (24/9)e^{4A}V^2$ for
homogeneous potential $V$. Now $V$ varies by $< 30$\% from inside the
bubble (false vacuum is $V(\phi_2)$) to the local maximum of the
potential (near $\phi \sim 1$; see Fig. 1). So, it is noteworthy that
$K$ in the regions where $A$ is maximum (inner parts of the bubble) is
subdominant to $K$ near the rim. The quadratic invariant is large near
the transition {\it just outside} the thin shell rim. It's divergence
is likely the signature of the inner false vacuum stretching or
detaching from the outer Minkowski space. The scalar exponentially
increases within a few time steps of size $\Delta t = 2.5 \times
10^{-4}$ and we halt the calculation at that point.

\newpage

\section{Numerical accuracy of solutions}

We describe three different checks on the accuracy of the
runs: (i) comparison of 1+1 and 2+1 results for a single
bubble, (ii) the fidelity of the initial transition profile,
and (iii) constraint violations.

\subsection{Numerical accuracy}

\begin{figure}[htbp]
\begin{center}
\begin{tabular}{ll}
\includegraphics[width=7cm]{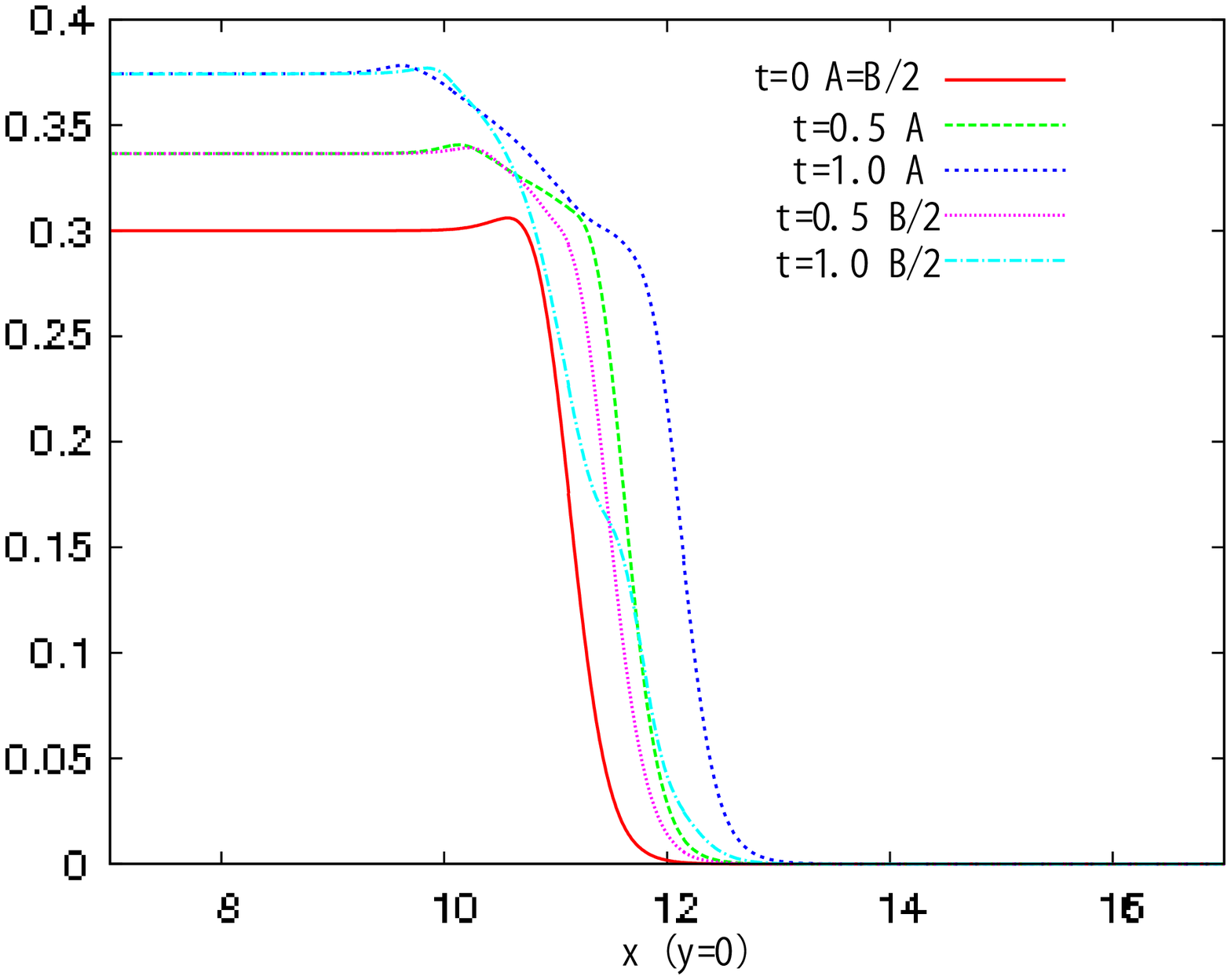} &
\hspace{1cm}
\includegraphics[width=7cm]{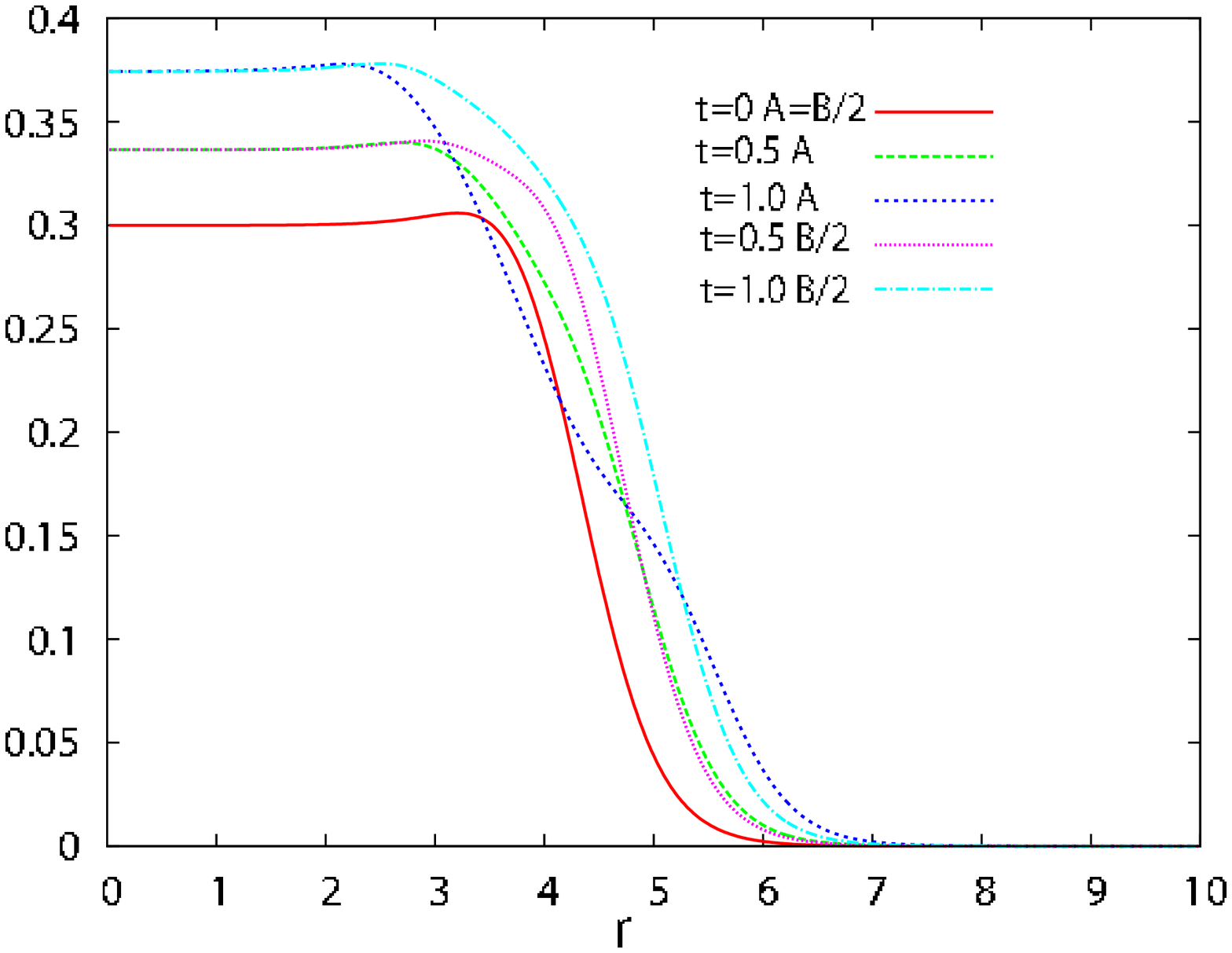}
\end{tabular}
\end{center}
\caption{The metric functions $A$ and $B$ are compared for one
  isolated bubble's evolution (snapshots at $t=0$, $0.5$ and $1$)
  calculated in two different coordinate systems.  The answers should
  be identical so the differences are an indication of the intrinsic
  numerical errors.  On the left the 2+1 Cartesian system $(x,y,t)$
  with bubble center at $(x,y)=(7,0)$ is used; on the right the 1+1
  cylinder system $(r,t)$ with $r=\sqrt{(x-7)^2+y^2}$ and bubble
  center at $r=0$ is used. The initial conditions are identical. We
  expect the 1+1 results to be more accurate based on the density of
  the grid and we infer that the differences reflect the
  characteristic size of the errors in the 2+1 treatment.}
\label{fig-nonbstab}
\end{figure}

\begin{figure}[htbp]
\begin{center}
\includegraphics[width=7cm]{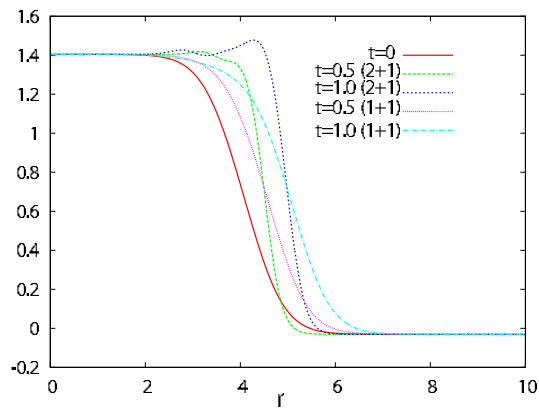}
\end{center}
\caption{Same as Fig. \ref{fig-nonbstab} except for
  $\phi(x,y,t)$ and $\phi(r,t)$ at selected times.
}
\label{fig-nonbstp}
\end{figure}

As a check on numerical accuracy, we compared the evolution of a
single isolated bubble calculated using two dimensional Cartesian
coordinates to one calculated with one dimensional polar coordinates.
Fig. \ref{fig-nonbstab} shows the result of $A$ and $B$ for a single
isolated bubble's evolution using the 2+1 ($(x,y,t)$) and 1+1
($(r,t)$) systems. The answers should be identical so the observed
differences are indicative of the size of numerical error. The 1+1
system has $2000$ radial zones while the 2+1 system has $10^6$
zones. The 1+1 system has effectively 5 times more points per given
area than the 2+1 system so we we anticipate the 1+1 results are more
accurate.\footnote{For the 1+1 system we repeated the calculations
  with $1000$, $1500$ and $2000$ points and observed effectively
  identical results for the latter two cases.}  The results are
approximately the same with the biggest notable difference to be found
in the quantity $B-2A$ at the bubble rim. Recall that the ratio of the
length in the z-direction to that in the r-direction is $e^{2(B-2A)}$
which is $>1$ in the 1+1 case but $<1$ in the 2+1 case. These
differences are indicative of the intrinsic errors in the evaluation
of the metric coefficients.
Fig. \ref{fig-nonbstp} compares the field $\phi$ for the 2+1 and 1+1
simulations. The 2+1 simulation apparently has a steeper transition
than the 1+1 simulation. The 2+1 calculation also appears to have some
oscillations in inner bubble edge that are not present in the 1+1
case. The bubble shears in both treatments. The inner region tracks
the thin wall limit and outer region moves outward in coordinates
position, with half-maximum $\phi$ at nearly the same point.

\subsection{Initial bubble profile}

\begin{figure}[htbp]
\begin{center}
\begin{tabular}{ll}
\includegraphics[width=6cm,angle=-90]{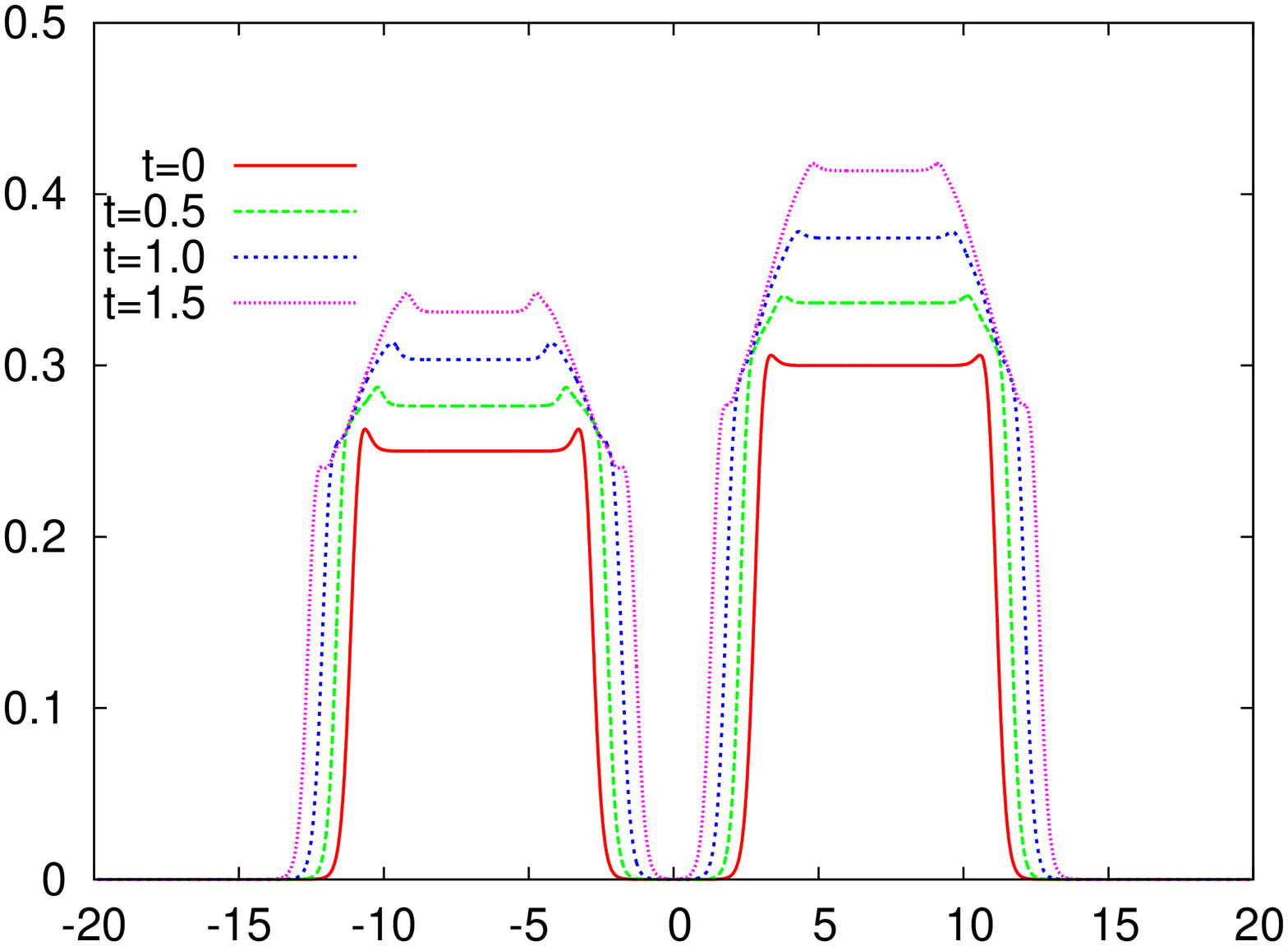} &
\hspace{0.5cm}
\includegraphics[width=6cm,angle=-90]{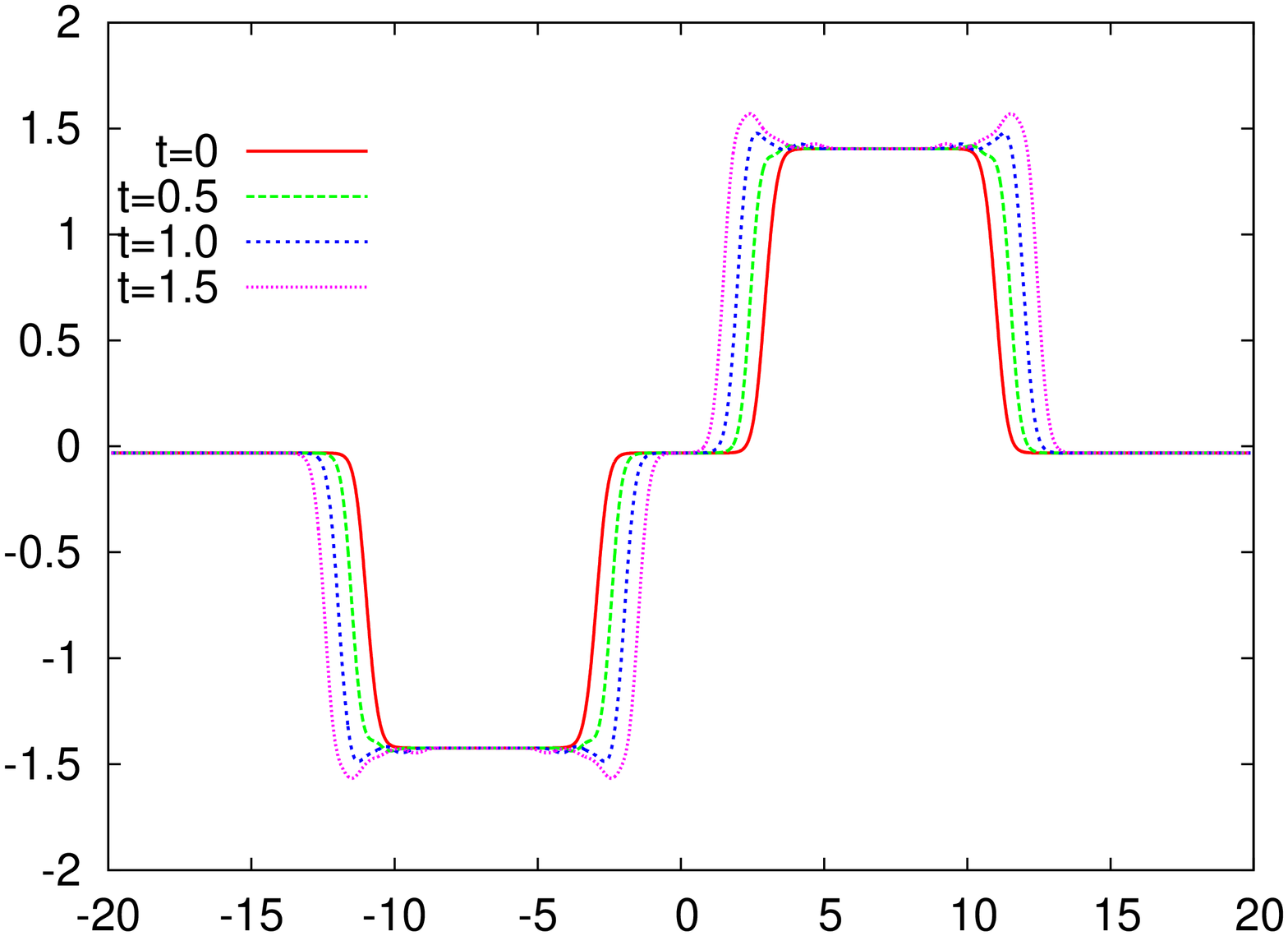}
\end{tabular}
\end{center}
\caption{The metric function $A$ and scalar field $\phi$ for spacetime
  with two evolving, isolated bubbles. Same as Fig. \ref{fig-nonbst1}
  but with $b_0=0.01$.}
\label{fig-nonbst2}
\end{figure}

The Hubble constants are $H_{1}=\sqrt{b_0} 0.520$ and $H_2= \sqrt{b_0}
0.716$. In fig. \ref{fig-nonbst2}, we change $b_0=1$ to $b_0=0.01$
and observe, as expected, that the Hubble expansion is about $1/10$-th
of the previous case. We will typically use $b_0 < 1$ to limit the
extent of exponential growth during the collisions we study. With the
smaller $b_0$ an profile change becomes apparent, the bubble
transition develops a horn-like structure near the rim as the field
relaxes.  This is related to the fact that the initial bubble profile
is approximate as described above: we start with the field $\phi$ held
fixed throughout the finite volume of the bubble and at the edges of
the box. It would be more realistic to impose zero radial field
derivative at the bubble center and let the field vary everywhere
between the center and the edges of the box. That is essentially what
happens once the simulation begins: the field profile relaxes as it
climbs the barrier separating the two minima. As the potential energy
decreases the field kinetic and gradient energy contributions become
relatively larger and their effect at the rim is more pronounced.

\subsection{Constraints}

\begin{figure}[htbp]
\begin{center}
\begin{tabular}{ll}
\includegraphics[width=6cm]{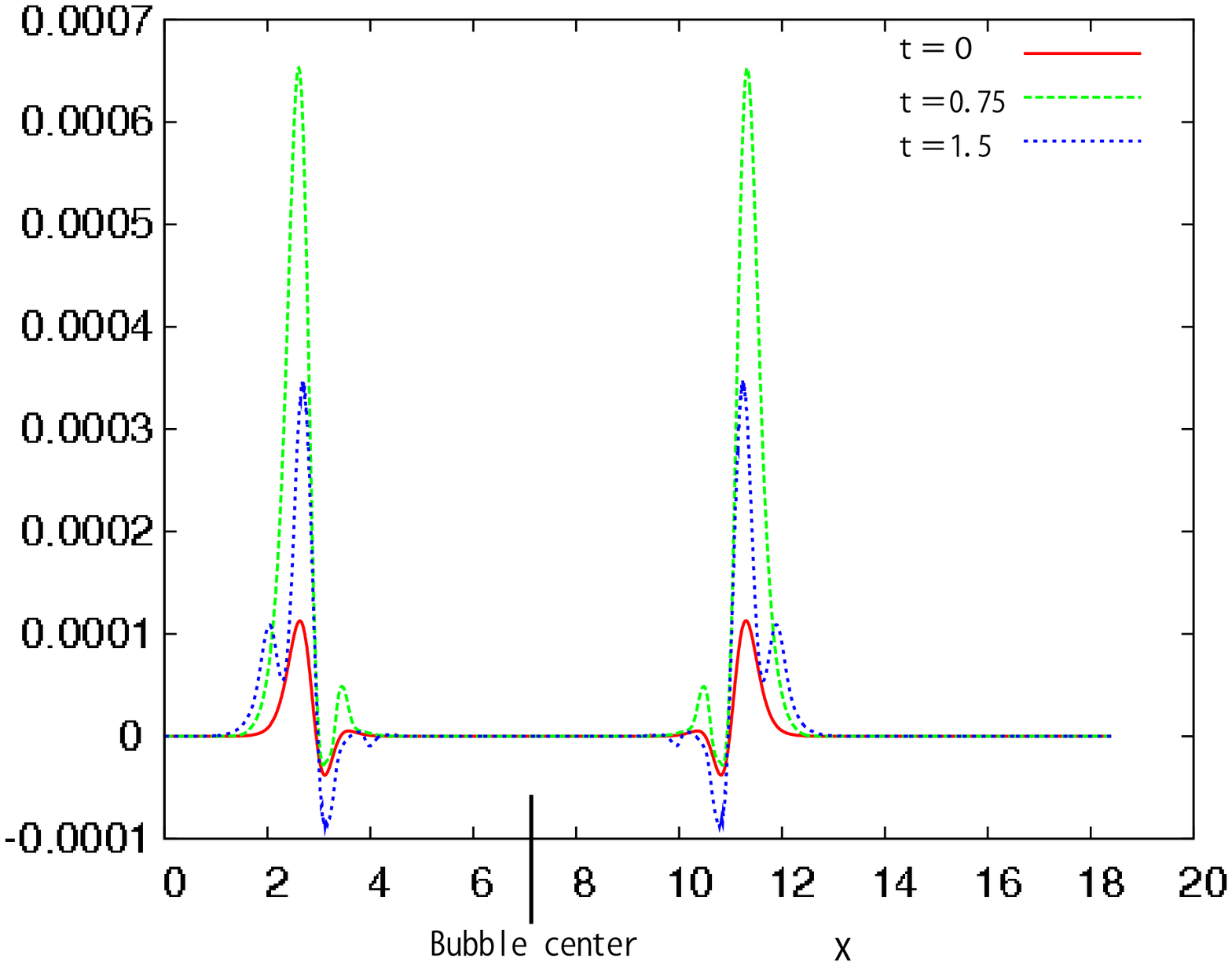} &
\hspace{0.5cm}
\includegraphics[width=6cm]{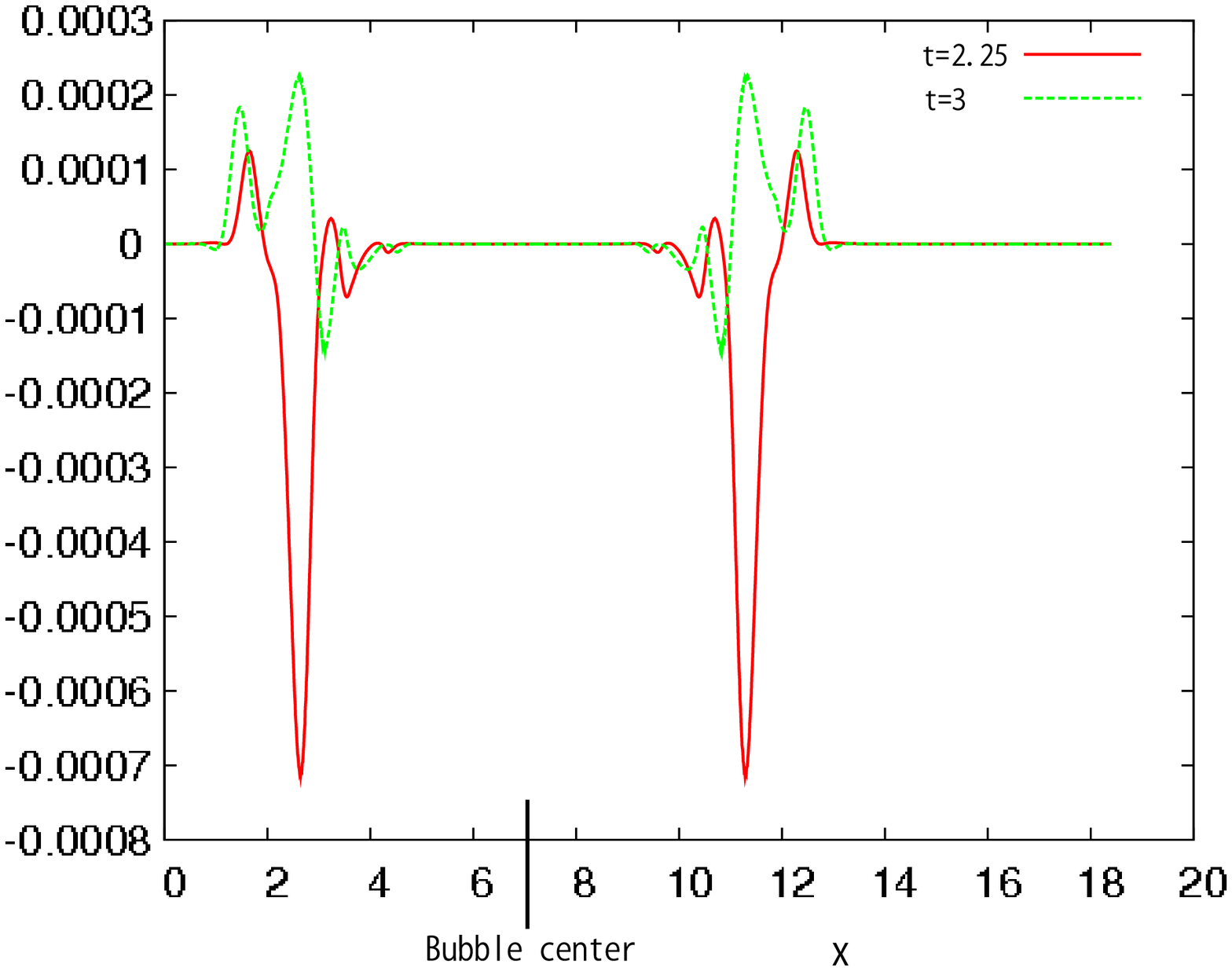}
\end{tabular}
\end{center}
\caption{The constraint violation in eqn. \ref{const1} at
  three early (left) and two later (right) times.}
\label{constraints}
\end{figure}

We have also checked the development of constraint violations
which is closely related to the accuracy with which we can
solve the equations of motion as well as our small deficit
angle approximation. For the effectively isolated
bubble simulations (illustrated in figs. \ref{fig-nonbst3},
\ref{fig-nonbst1},
\ref{forces},
\ref{isolated-bubble},
\ref{fig-RRx},
\ref{fig-RRap})
we find eqn. \ref{const1} is satisfied to a relative
error of $< 10^{-3}$ (the residual result for eq. \ref{const1}
divided by the sum of
the absolute value of each separate term) until the
Kretschmann scalar diverges near $t = 3$. The relative error
is illustrated at 5 snapshots in fig. \ref{constraints}.
The constraint violation is not monotonically increasing
although it does tend to grow with time.

\section{Initial conditions for moving bubbles}

We outline our general method to construct initial data for an
arbitrary collision between two moving, hitherto non-interacting,
bubbles.  We can transform an isolated bubble solution from frame $S$
to another frame $S'$ when one frame moves with constant velocity with
respect to the other. This is a simple Lorentz transformation. In our
problem $S$ is the frame in which a bubble has nucleated (bubble
center is at rest) and $S'$ is the collision frame. In our setup with
two bubbles the left and right bubbles do not interact initially
because they are separated by Minkowski space ($A=B=\phi=0$) and the
rims expand at less than the speed of light. We boost each bubble
separately and stitch two halves at $x'=0$ together to create initial
conditions that will eventually give rise to a collision.

\begin{figure}[htbp]
\begin{center}
\includegraphics[width=6cm]{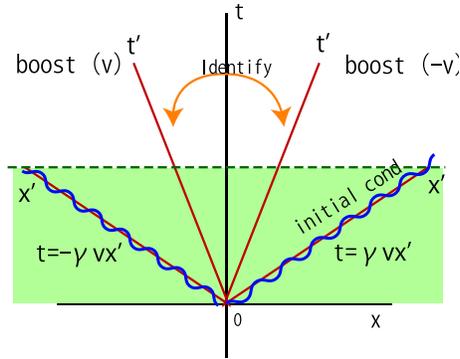} 
\end{center}
\caption{A schematic picture in frame $S$ illustrating times of
  simultaneity for two different boosted frames $S'$.  To the right of
  the origin the boost is $-\up<0$; to the left it is $\up > 0$.  We
  solve the dynamical equation for $A$, $B$ and $\phi$ for isolated
  bubbles in frame $S$ throughout the spacetime region shaded
  green. We boost these solutions to frame $S'$ (separately on the
  right and left sides) and infer initial conditions in the collision
  frame.}
\label{fig-collide}
\end{figure}

Let the origin of $S$ move with velocity $W'$ in $S'$. We want the
bubble at position $x=x_1<0$ to move towards the origin in $S'$; we
take $W' > 0$ and write the desired motion for bubble 1 as $W' = W'_1
= \upsilon$.  Likewise for the bubble at position $x=x_2>0$ we want
$W' < 0$. Here, we write $W' = W'_2 = -\upsilon$.  The Lorentz
transformations are
\begin{align}
t=\gamma (t'-W' x')\,,~~x=\gamma (x'-W' t')\,,~~z=z'\,,
\label{eq-Lorentz}
\end{align}
for velocity $W'=\pm \upsilon$ and $\gamma=1/\sqrt{1-\upsilon^2}$.  We
need not adopt $W'_1 = -W'_2$ as long as the transformed bubbles
remain well-separated at $t'=0$.  Note that at time $t'=0$ we have
$t=-\gamma W' x' = -x W' > 0$ because $x_1<0$ has $W'_1>0$ and $x_2>0$
has $W'_2< 0$. The initial condition slice at $t'=0$ is shown in
Fig. \ref{fig-collide}.

The form of the metric (\ref{eq-metric}) is invariant under this
transformation. The new scalar function $A'$ is $A$ evaluated at the
transformed coordinates, i.e. $A'(t',x',y')=A(t,x,y)$ where $t$, $x$
and $y$ are understood to be functions of $t'$, $x'$ and $y'$
according to the Lorentz transformation with the appropriate boost
$W'=\pm \upsilon$.  Omitting the independent variables we have
schematically
\begin{align}
&A'|_{t'=0}=A |_{t=-\gamma W' x'}\,,\\
&A'_{t'}|_{t'=0}= \gamma A_t |_{t=-\gamma W' x'}-\gamma W' A_x |_{t=-\gamma W' x'} \,.
\end{align}
The initial data is derived from two slices of evolved bubble solutions in the
original coordinate system where $t=-x W' > 0$ as illustrated in
Fig. \ref{fig-collide}. We infer $\phi'$ and $\phi'_{t'}$ at $t'=0$ in an
identical fashion.

Once we have found the metric and the scalar field (values and time
derivatives), we integrate equations (\ref{dyn1}) and (\ref{dyn2}) in
the new coordinates. Henceforth, we drop the explicit use of primes
and simply refer to the coordinates in the collision frame as $t$, $x$
and $y$.

\begin{figure}[htbp]
\begin{center}
\begin{tabular}{ll}
\includegraphics[width=6cm]{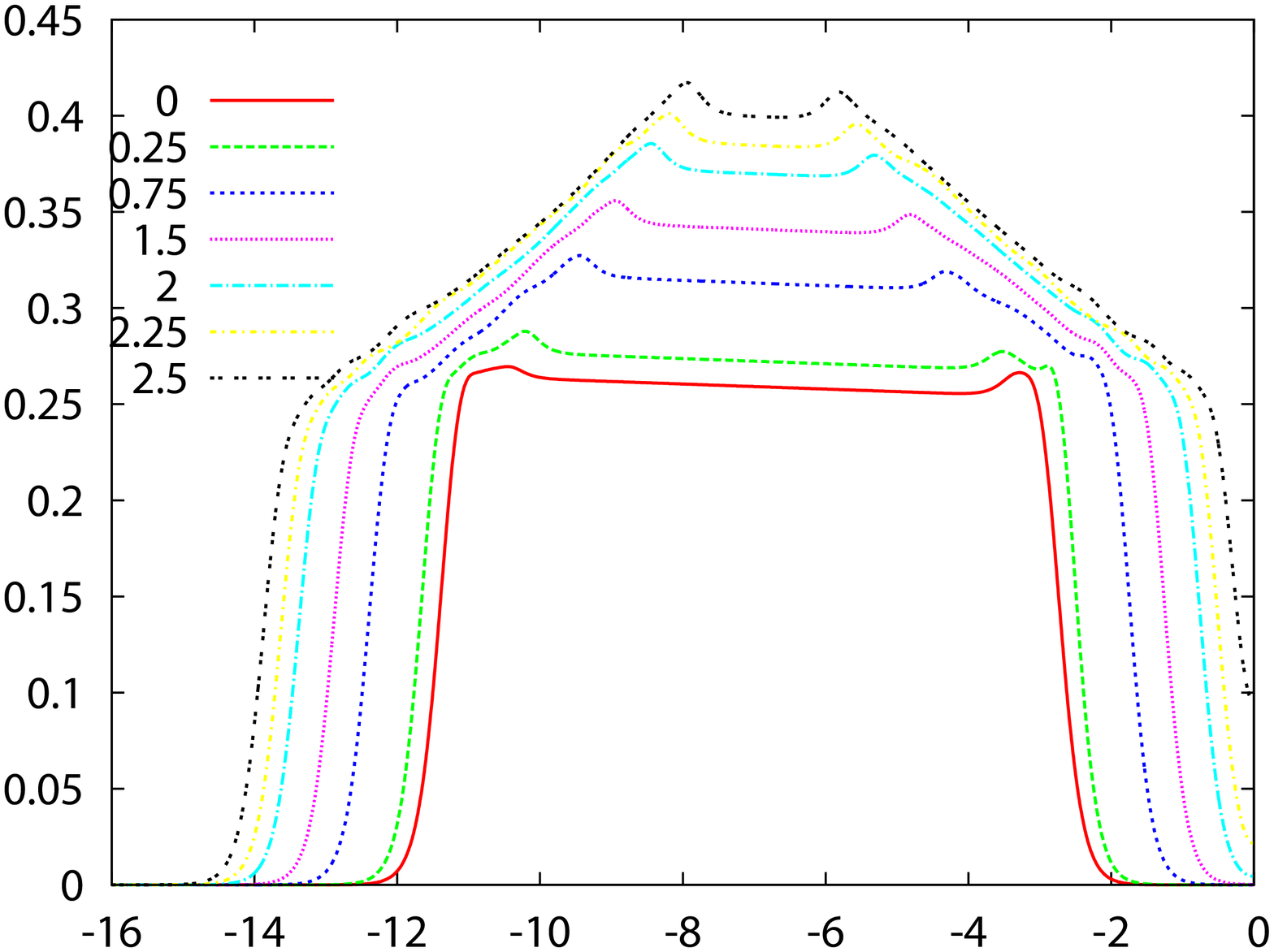} &
\hspace{1cm}
\includegraphics[width=6cm]{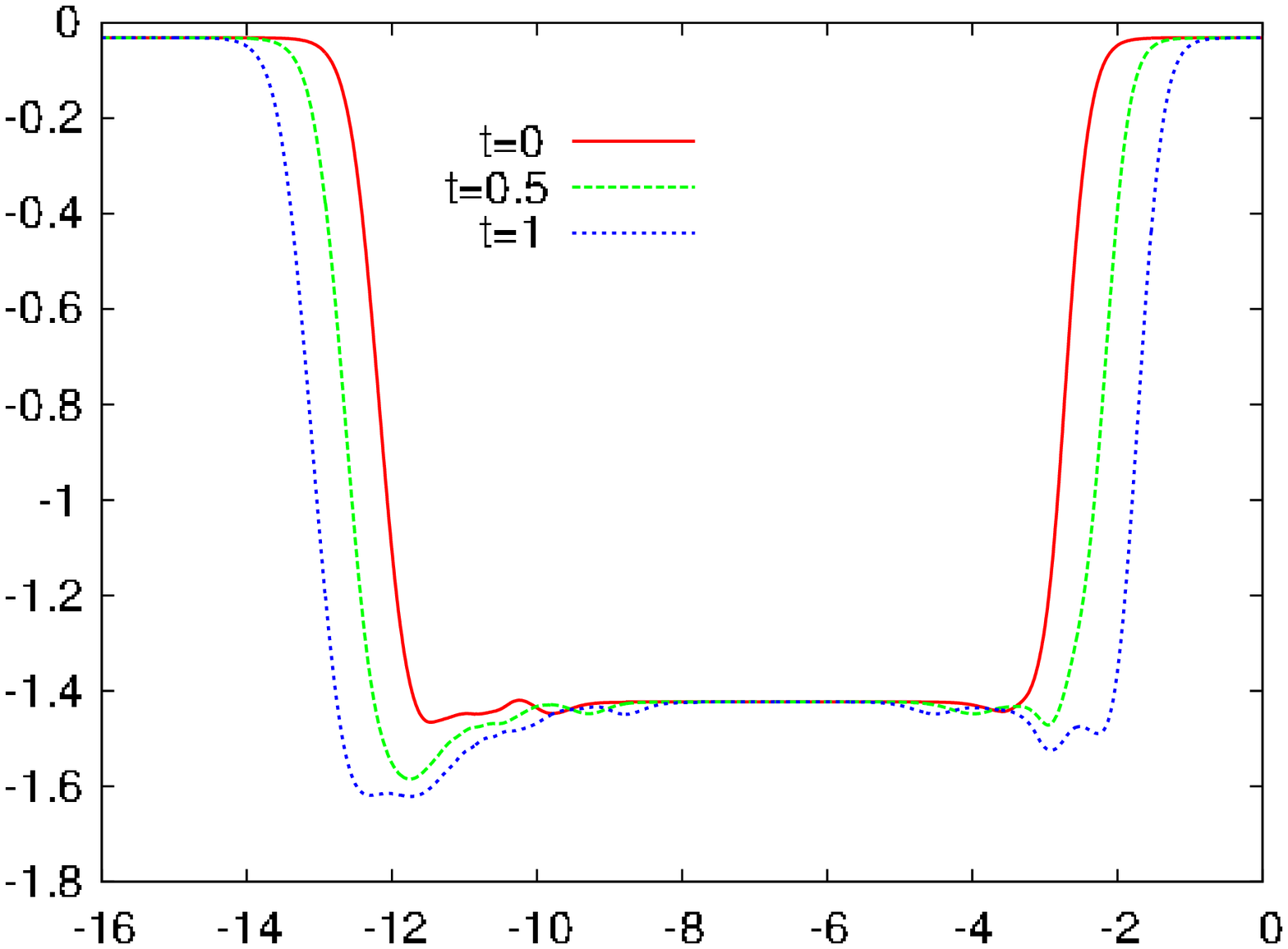}
\end{tabular}
\end{center}
\caption{We plot one boosted bubble solution in the collision frame
  along $y=0$ at three times $t=0$ (red), $t=0.5$ (green) and $t=1$ (blue).
  $A$ ($\phi$) is shown in left (right) figure. The
  boost is $\up=0.1$ in this region. In the left figure the highest
  point of $A$ lies close to the inner edge of the false
  vacuum. It moves upward and rightward with time,
  tracking the motion of the thin shell.  In the accompanying
  right hand figure we see the motion of the outer, transition
  regions. Both walls expand (time snapshots from red to blue) at
  close to $v={\cal O}(1)$, much larger than that of the boost
  itself.}
\label{fig-bst}
\end{figure}

We advance the solution with small time steps $\Delta t=10^{-4}$ using an
explicit, 4th-order Runge-Kutta method to update $A$, $B$ and
$\phi$. The spatial derivatives are formed by finite difference
approximations.  Fig. \ref{fig-bst} shows the evolution of one bubble
in the collision frame. The plot on the right
  shows that the wall moves at nearly the speed of light to the
  left at small $x$. In the collision frame the bubble rim moves
with velocity
\begin{equation}
{\dot R}_{coll}={\frac{dx}{dt} +\frac{{\dot R}_{nuc}}{1+\frac{dx}{dt}{\dot R}_{nuc}}},
\end{equation}
where the boost velocity
$\frac{dx}{dt}=\pm\upsilon$ and ${\dot R}_{nuc}$ is the rim velocity in
nucleation frame. With $\frac{dx}{dt}=0.1$ and ${\dot R}_{nuc}=0.9$ the total
velocity ${\dot R}_{coll}=0.92$. In this numerical example the total velocity
of approach is dominated by the bubble expansion.

\newpage

\section{Bubbles collide and begin to interact}
\begin{figure}[htbp]
\begin{center}
\begin{tabular}{ll}
\includegraphics[width=8.5cm]{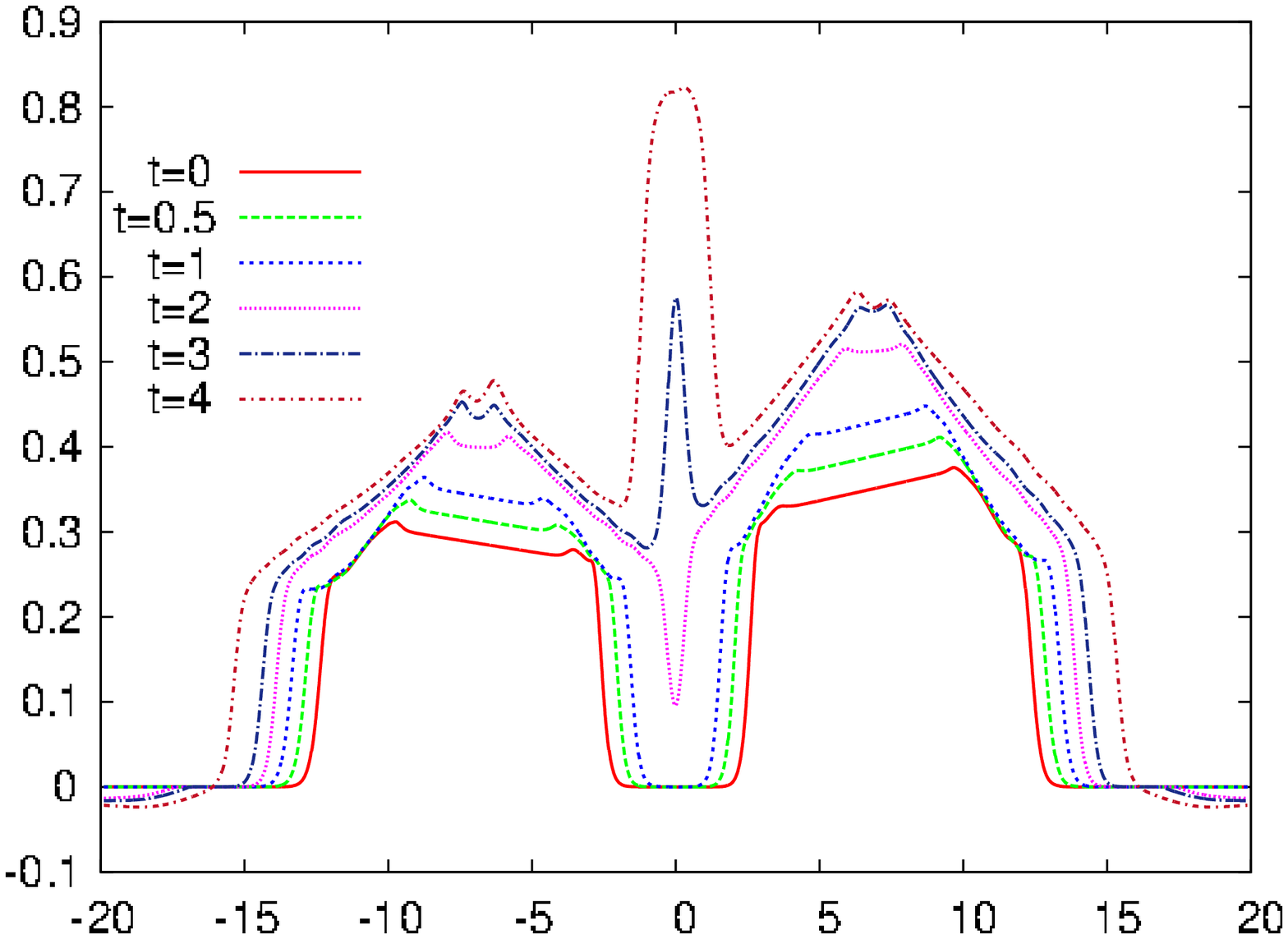} &
\hspace{0.5cm}
\includegraphics[width=8cm]{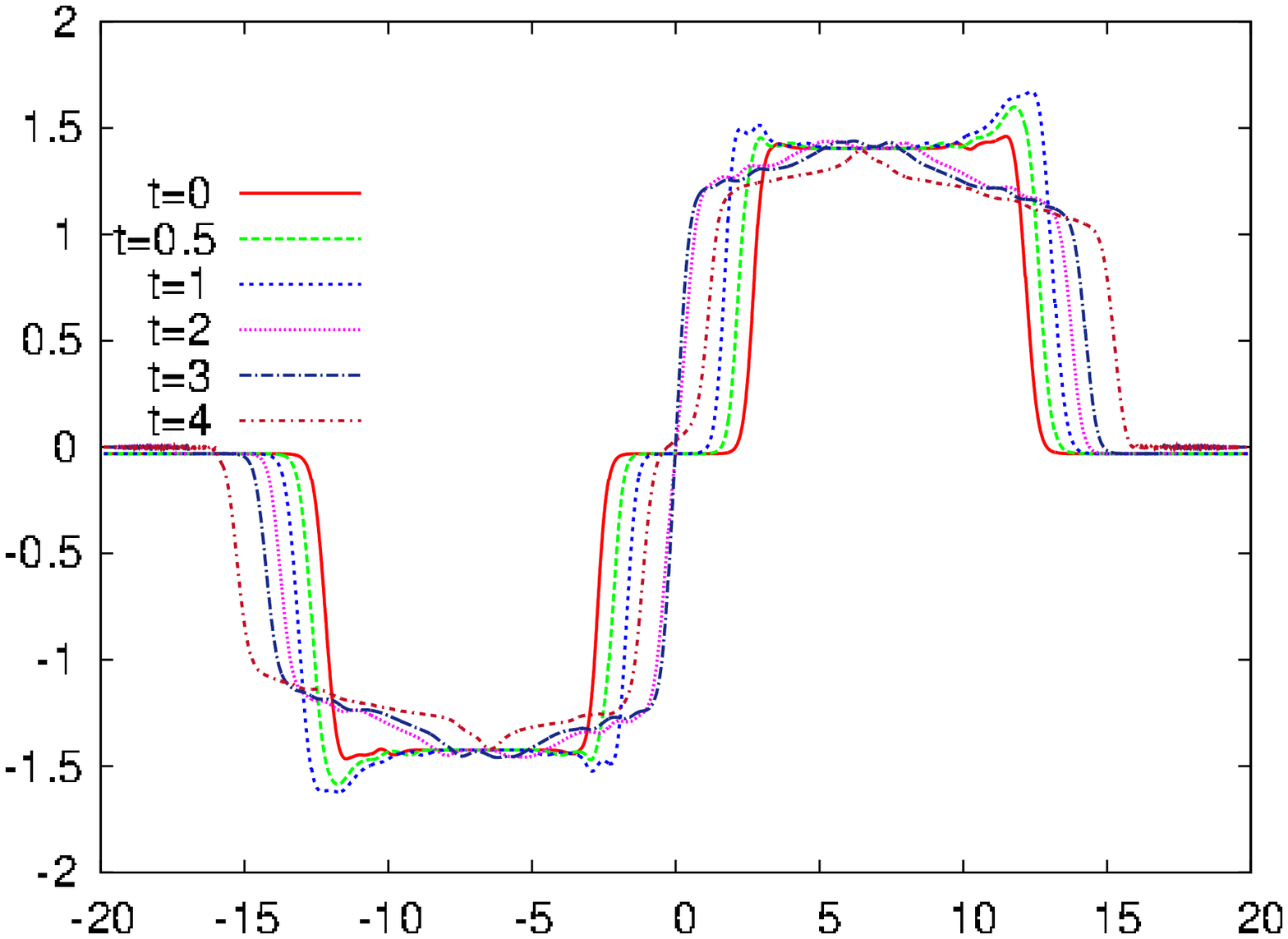}
\end{tabular}
\end{center}
\caption{Snapshots in time leading up to the collision of two bubbles
  in the collision frame.  On the left $A(t,x,y=0)$ and on the right
  $\phi(t,x,y=0)$ where the plane of symmetry is $y=0$. As the bubbles
  begin to connect the region between them experiences a greater rate
  of expansion than either individual bubble. Estimating from the left
  hand figure we have $\Delta A(x=0)\simeq 0.5$ during $\delta t = 3 -
  2 \simeq 1$, or $H(x=0) \sim 0.5$. For comparison, $H_1=0.05$ and
  $H_2=0.07$ in the nucleation frames and not too different in the
  collision frame. The right hand picture illustrates that the field
  which must stretch from $\phi_1$ to $\phi_2$ does so over a
  shrinking $x$ coordinate range.}
\label{fig-coll2}
\end{figure}

Fig. \ref{fig-coll2} shows the collision of two bubbles.  The left
hand bubble is the bubble whose isolated evolution was described in
the previous section.  The figure shows the metric function $A$ and
the field $\phi$ along a slice through the bubble centers.

\begin{figure}[htbp]
\begin{center}
\begin{tabular}{ll}
\includegraphics[width=9.5cm]{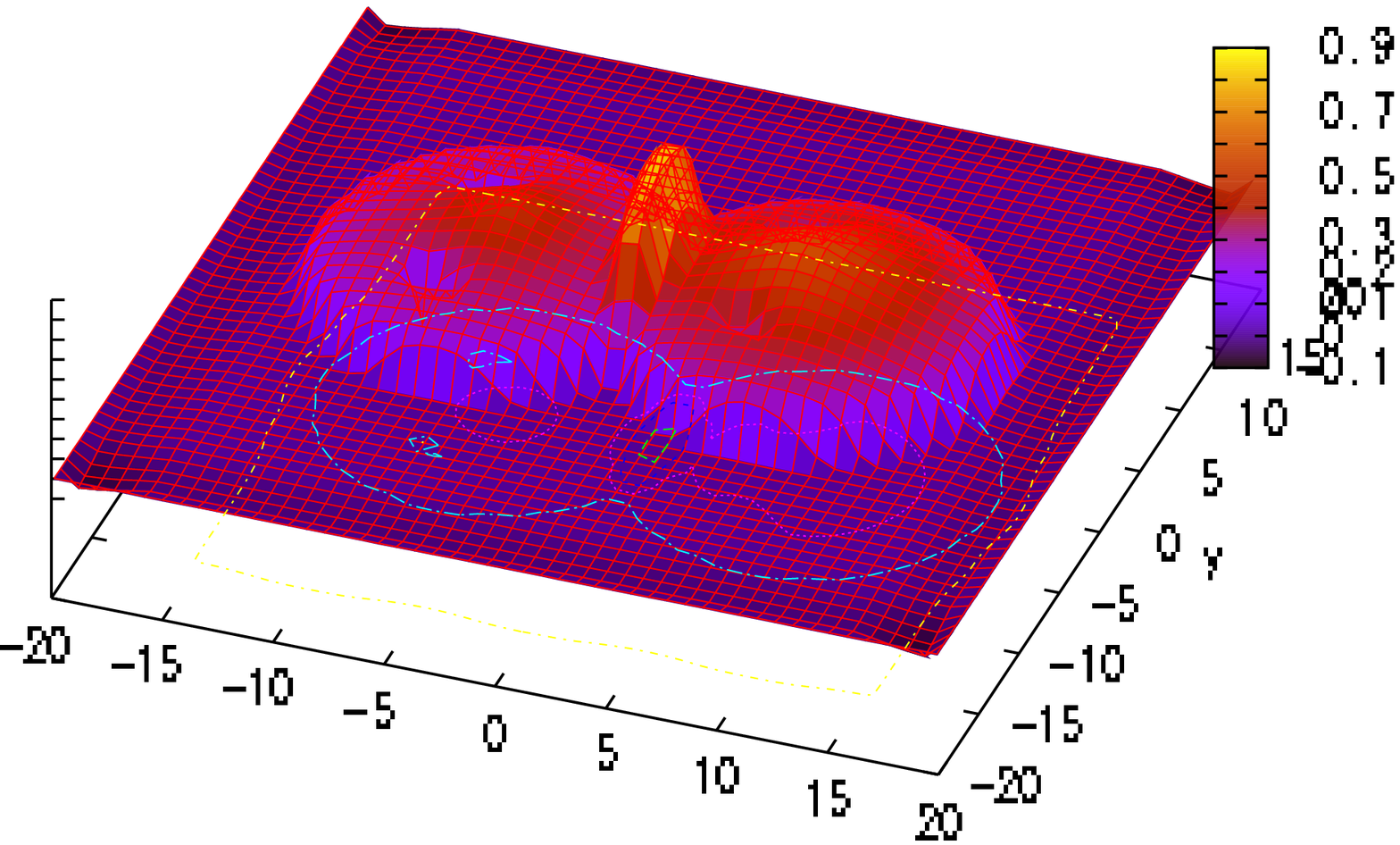} &
\hspace{0.5cm}
\includegraphics[width=8cm]{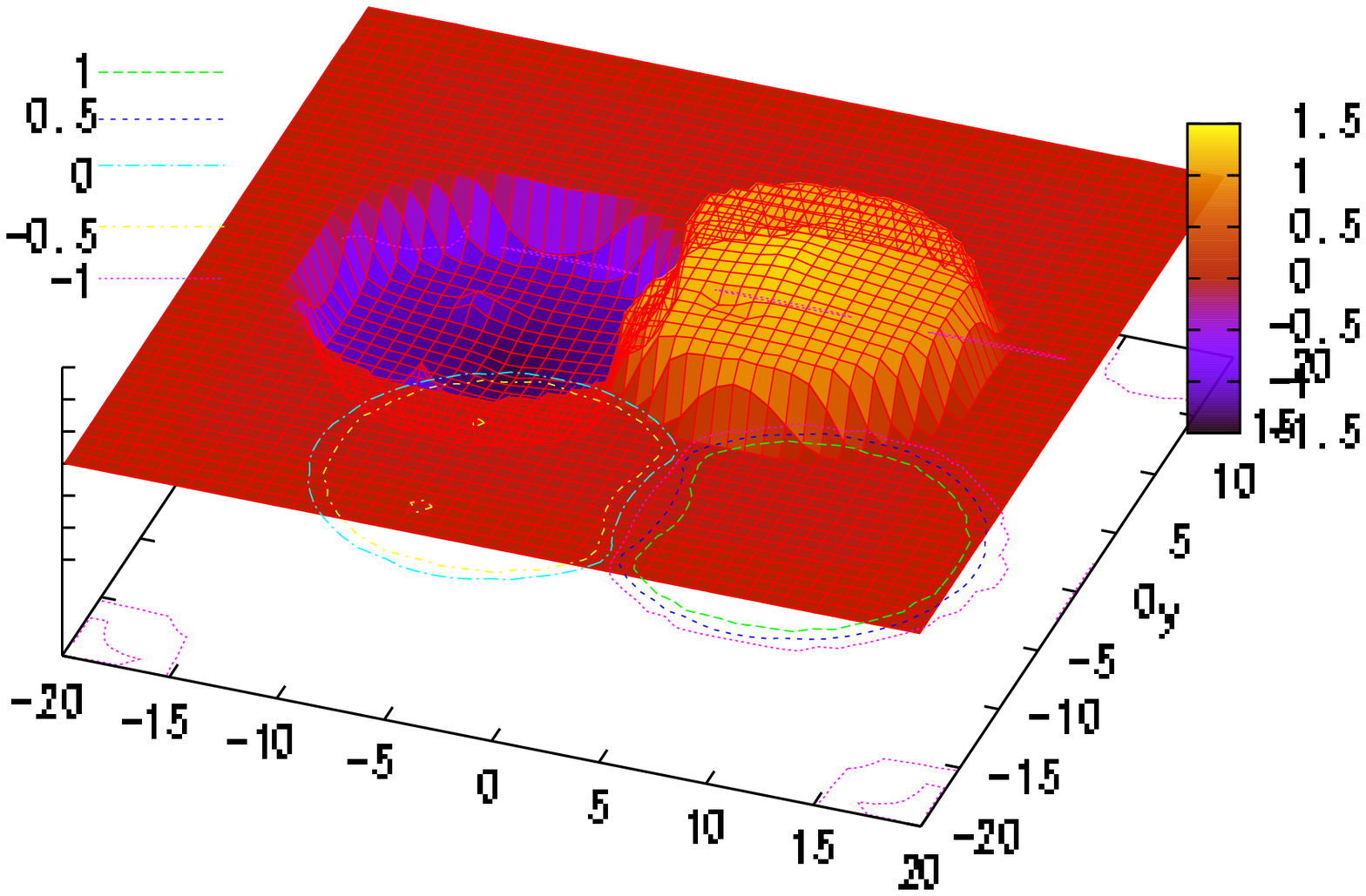}
\end{tabular}
\end{center}
\caption{A surface plot of $A$ (left) and $\phi$ (right) at $t=4$ when
  two bubbles begin to collide. These show the region of the
  collision. The colored lines below the surface plot are contours for selected
  values of $A$ and $\phi$. 
}
\label{fig-coll1}
\end{figure}

Fig. \ref{fig-coll1} displays the corresponding surface plot for $A$
and $\phi$ at $t=4$.  The terms in the metric dependent on $A$ and $B$
grow exponentially. Note that different regions acquire different
effective cosmological constants and that the expansion in the region
of interaction is much larger than within the unperturbed bubbles.

The FLRW equation for maximal symmetry (eq. \ref{const-H} and the
approximation $B=2A$) implies that the conformal Hubble expansion
(\ref{Hubble-eq}) depends upon the gradient of scalar field, $\phi_x$,
its time derivative, $\phi_t$ and a gradient of the metric, $A_x$:
\begin{equation}
6H^2(t,x,y)=4(A_{xx}+A_{yy}) +2(A_x^2+A_y^2)+\phi_t^2+\phi_x^2+\phi_y^2+2e^{2A}V(\phi)\,,\label{const-H1}
\end{equation}
The dominant contribution on the right hand side may be traced to the
large collision induced spatial gradients in the field $\phi$.
Numerically at $t=3.0$ we have $\phi_x^2\sim 8$, $4 A_{xx}\sim -3$ and
$2A_x^2 \sim \phi_t^2+2 e^{2A}V \sim 1$ at the origin. They lead to
$H\simeq O(1)$.  For the assumed geometry the scalar field must
connect two different local minima $\phi_1$ and $\phi_2$ that lie on
opposite sides of global minima at $\phi_0$ (specifically, $\phi_1 <
\phi_0 < \phi_2$) across a distance that shrinks as the two bubble
rims approach.  Fig. \ref{fig-coll2} illustrates the phenomenon; see
the blue line in the left panel. The gradient in the collision region
exceeds that near an isolated bubble wall.

\newpage

\section{Bubble collision without boost}
\begin{figure}[htbp]
\begin{center}
\begin{tabular}{ll}
\includegraphics[width=8.5cm]{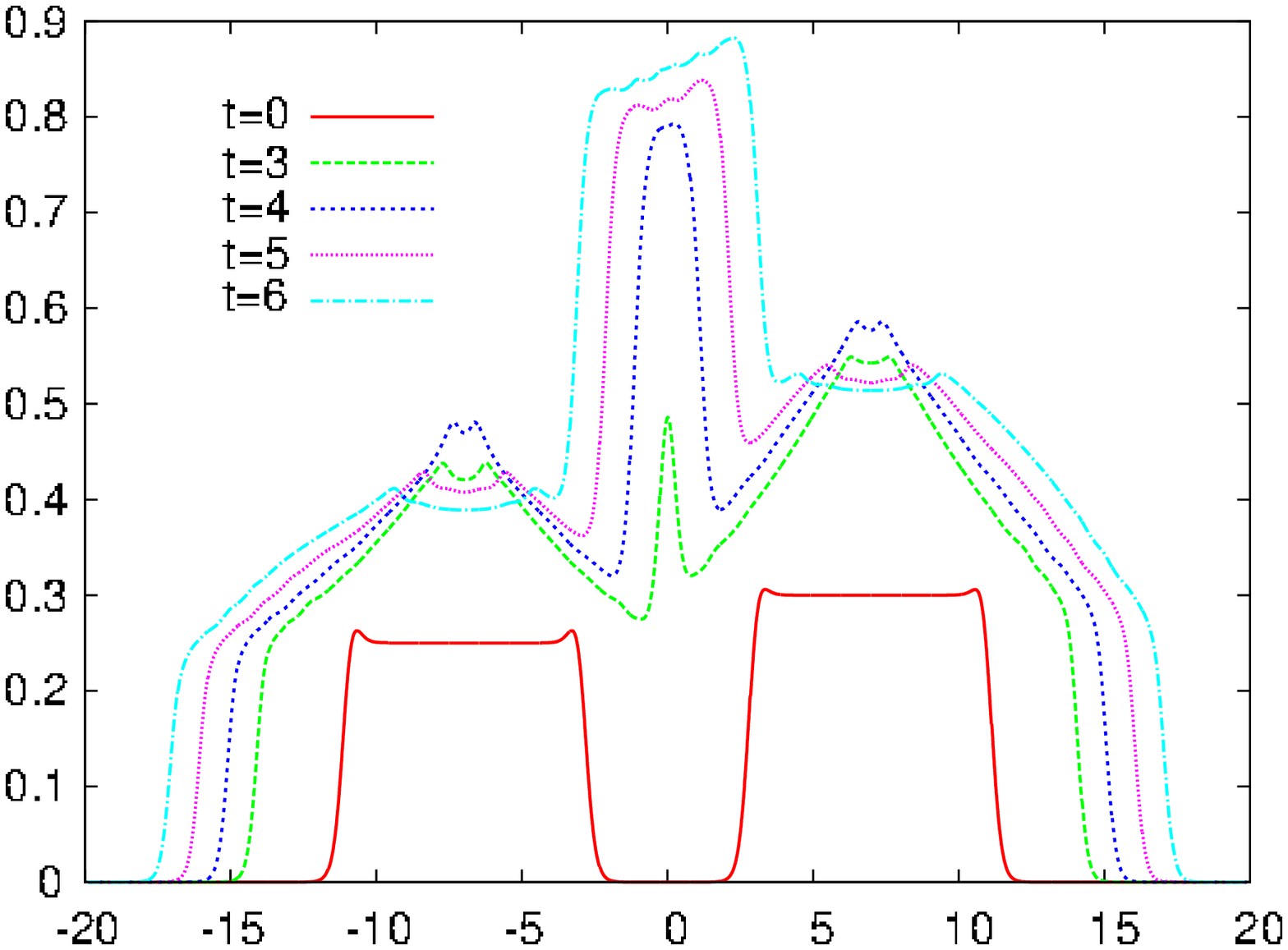} &
\hspace{0.5cm}
\includegraphics[width=8cm]{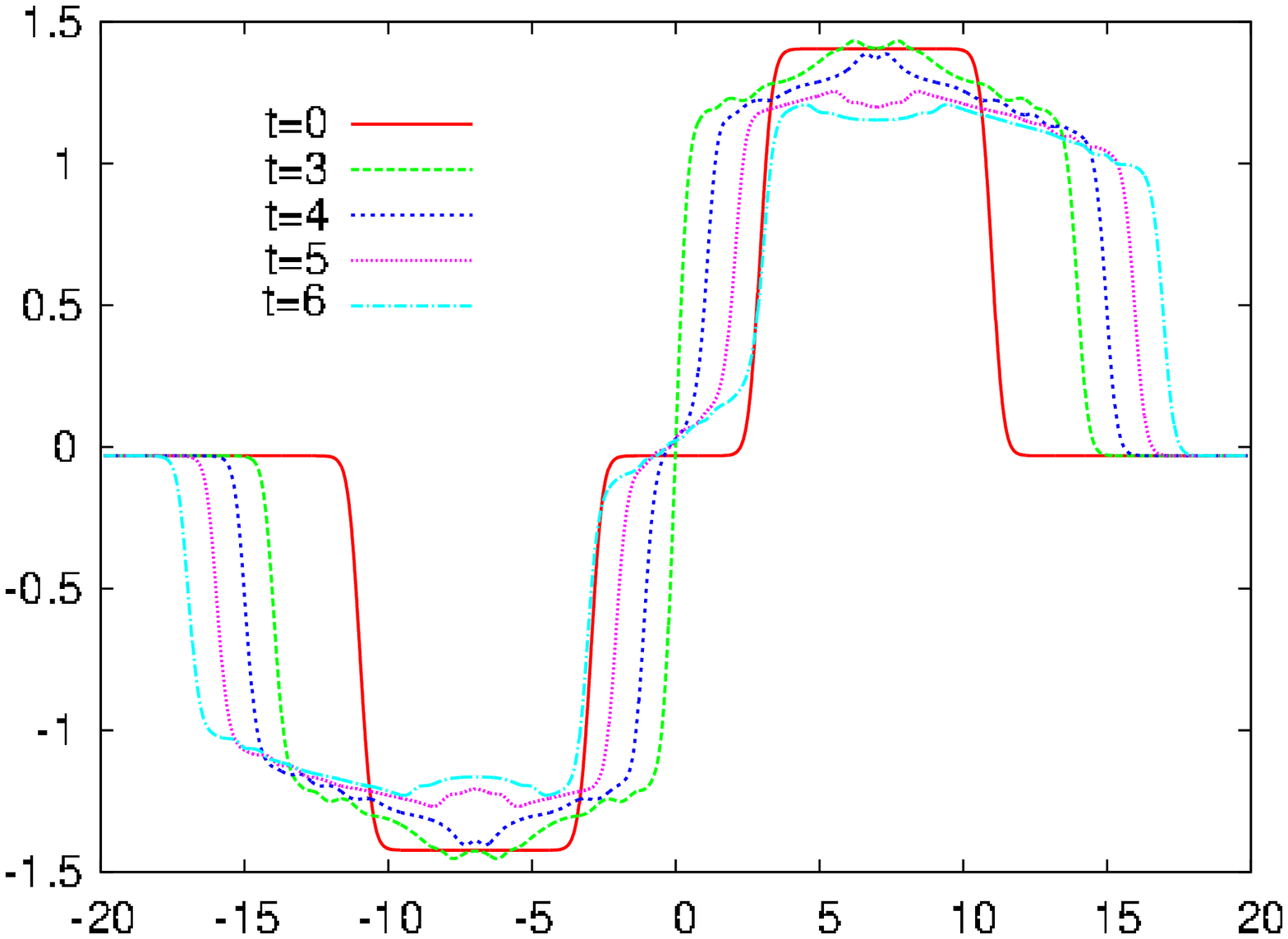}
\end{tabular}
\end{center}
\caption{Snapshots in time of $A$ (left) and $\phi$ (right) at $y=0$
for two colliding bubbles. The simulation stops at $t=6.8$.}
\label{fig-coll3}
\end{figure}

\begin{figure}[htbp]
\begin{center}
\begin{tabular}{ll}
\includegraphics[width=9.5cm]{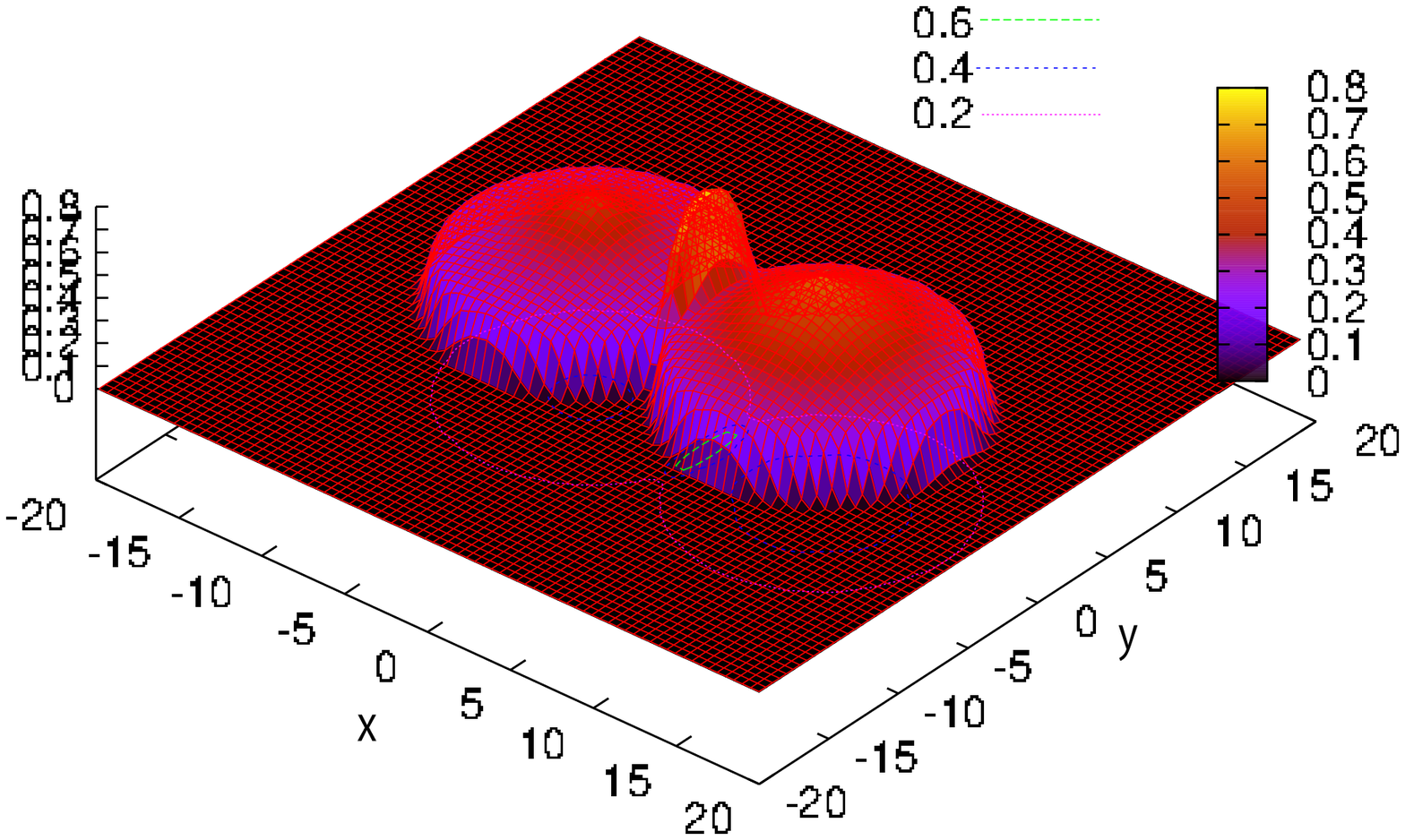} &
\hspace{0cm}
\includegraphics[width=8cm]{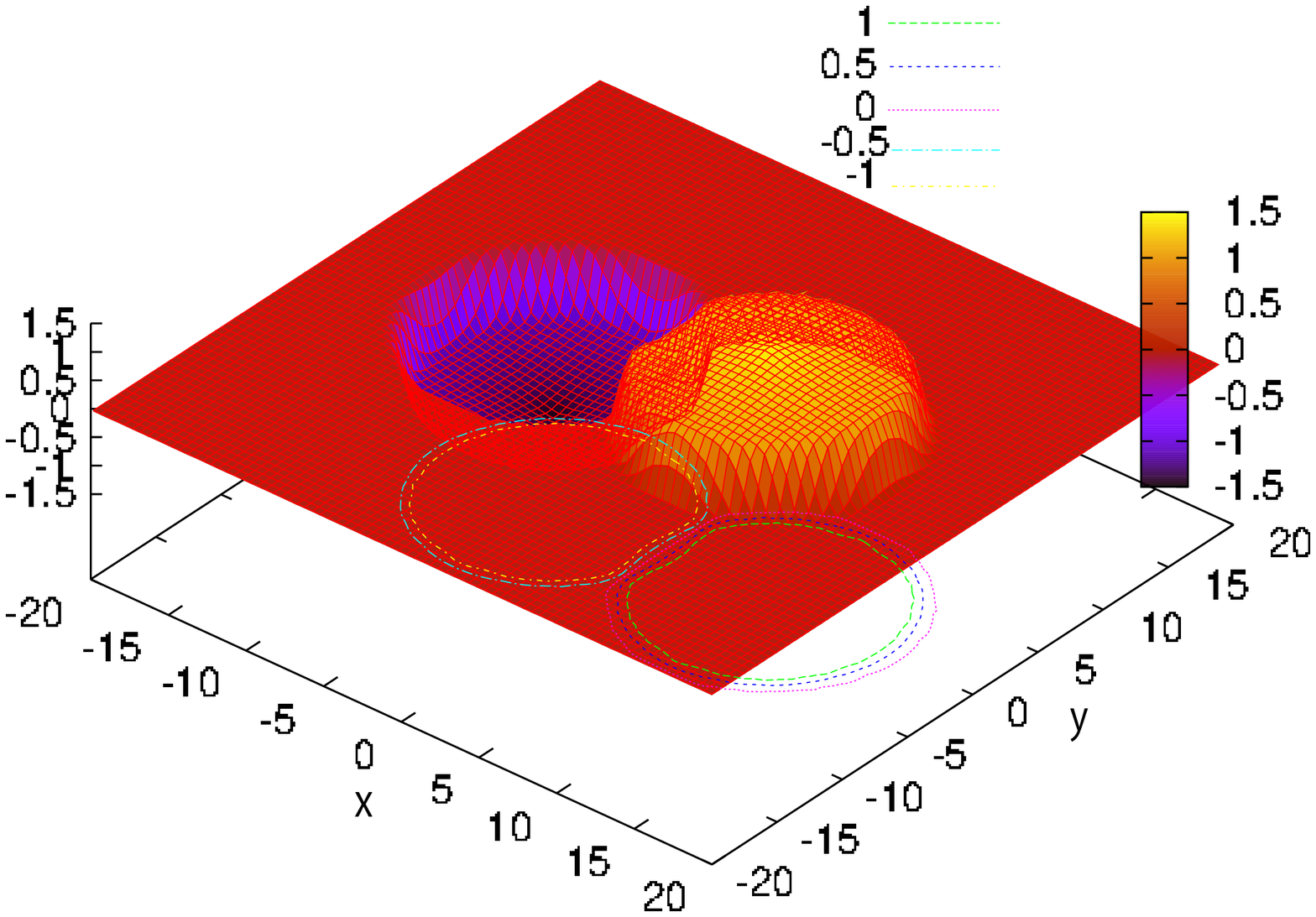}
\end{tabular}
\end{center}
\caption{Surface plot for 
$A$ (left) and $\phi$ (right) at $t=5.5$}.
\label{fig-coll4}
\end{figure}

\begin{figure}[htbp]
\begin{center}
\begin{tabular}{ll}
\includegraphics[width=8.5cm]{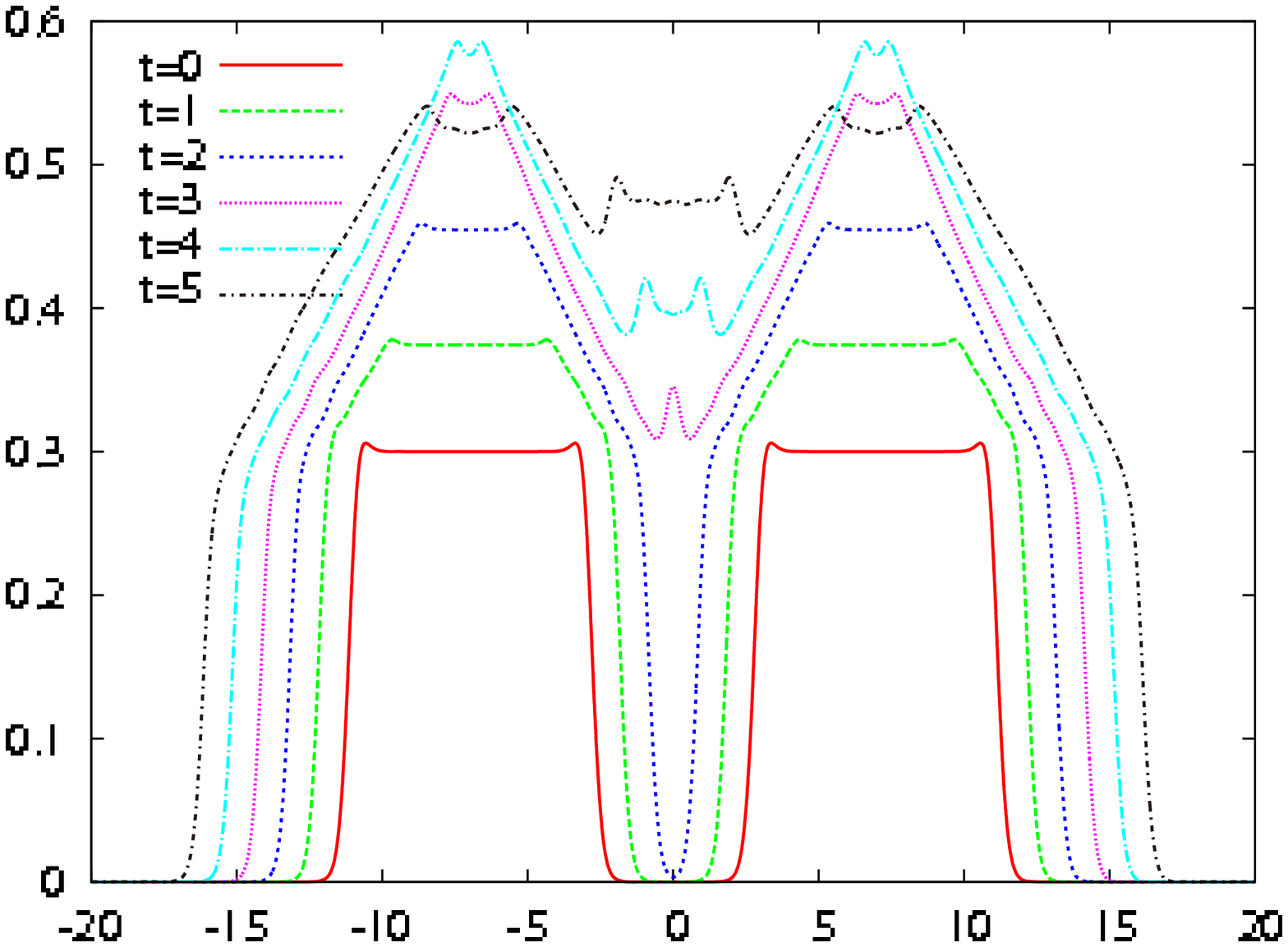} &
\hspace{0.5cm}
\includegraphics[width=8cm]{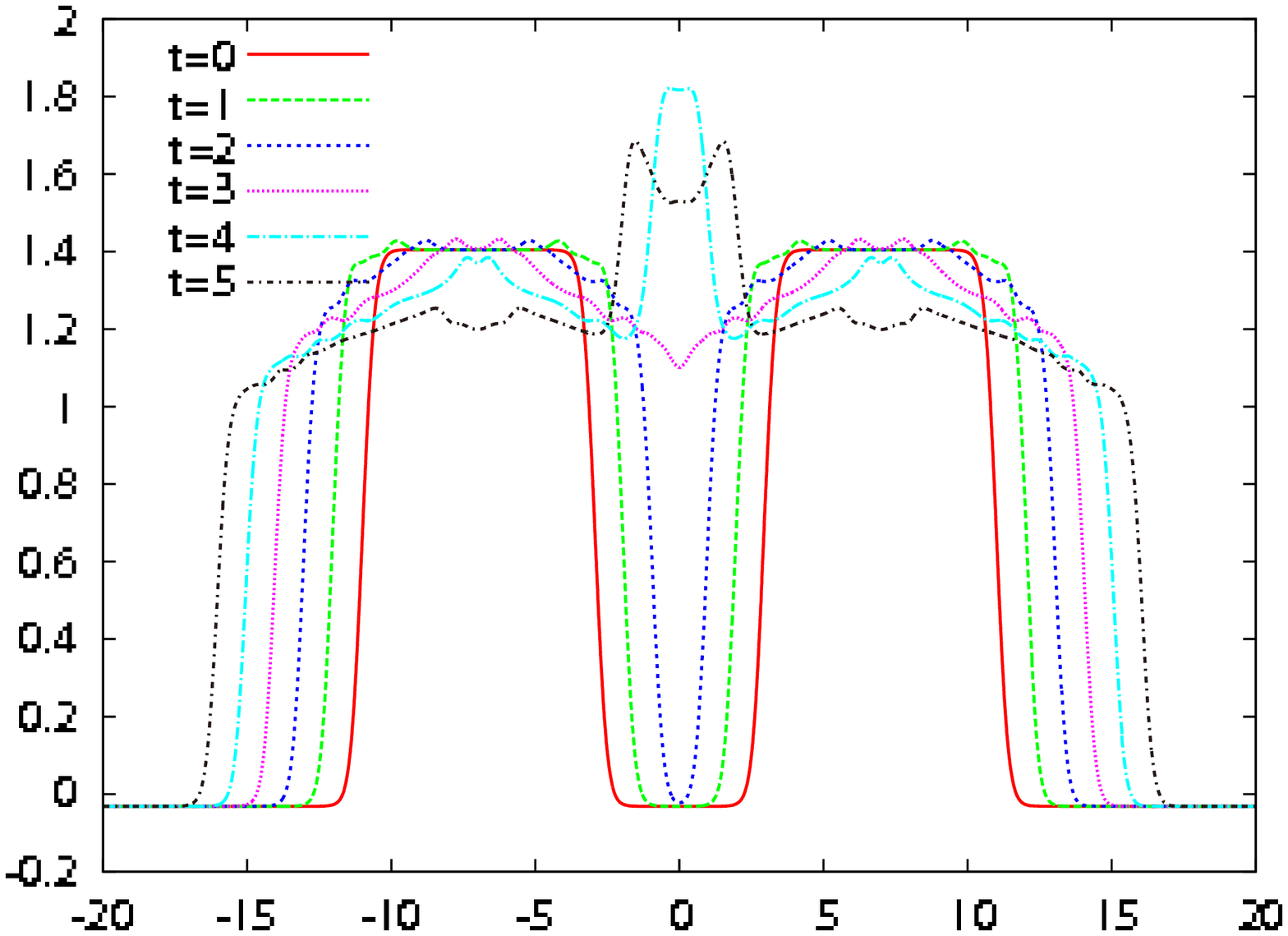}
\end{tabular}
\end{center}
\caption{We plot time evolution of $A$(left) and $\phi$(right) at $y=0$ 
until collision of two bubbles. The simulation stops at $t=5.5$.  
The scalar field exceeds $\phi=1$ (where $V(\phi)$ has an unstable
equilibrium) throughout most of the collision region. It is unclear which
parts will return to the global minimum and which parts to the local
metastable minimum.}
\label{fig-coll5}
\end{figure}

We have seen that the bubble speed in the collision frame ${\dot
  R}_{coll}$ depends upon the boost $dx/dt$, the velocity of the
nucleated bubble center in the collision frame, and ${\dot R}_{nuc}$,
the velocity with which the bubble rim moves in the nucleation frame.
If the bubble rim dominates then the distinction between nucleation
and collision frames is immaterial. Suppressing the boost makes some
physical features more readily apparent. From now on we begin with two
bubbles with stationary centers that interact when their rims meet
(Figs. \ref{fig-coll3}, \ref{fig-coll4} and \ref{fig-coll5}).

The features already identified in the collision frame are clearer in
Figs. \ref{fig-coll3} and \ref{fig-coll4}. The initial bubble rims are
identical but the vacua are distinct.  The region of interaction
experiences a larger rate of expansion than that of either original
bubble. The effective Hubble constant during interaction is an order
of magnitude larger than that of the individual, non-interacting
bubbles.  The growth of the metric and the potential in the central
region is a multi-step process: the kinetic energy of the bubble walls
is localized by the aligned 2+1 collision geometry and focused by
increasing spacetime curvature. Part of the field kinetic energy
$\phi_t^2/2$ is transformed into potential energy as the field moves
away from its global minimum. The field's evolution approaches a
classical turning point much like a particle moving towards a rising
potential, i.e. the field kinetic energy $\phi_t^2/2 \to 0$ and the
field potential energy $V$ becomes large everywhere except at the
field point where $V=0$. Since spacetime is not homogeneous the turn
around does not occur at a single, well-defined point of time.
Spatial gradients like $\phi_x^2$ are enhanced during the collapse and
contribute to the large, effective Hubble constant
  
Fig. \ref{fig-coll5} shows the case when the collision involves two
bubbles with the {\it same} false vacuum. The asymmetry apparent in
the previous case is suppressed. The kinetic energy of the colliding
walls plus the vacuum energy remains. The metric coefficient grows but
by a much smaller amount and $H\simeq 0.1$. The symmetry here is
important and the gradient term, $\phi_x^2$, is not as large in the
aftermath of the collision as it is in the asymmetric case. In the
asymmetric collision, it provided the dominant contribution to $H$.

\newpage

\section{Long time results}

An important question is the asymptotic outcome of the bubble
collision. Which parts of spacetime will remain in metastable local
minima and which parts will return to the global minimum?  There are
two issues which will limit the extent of the runs we can perform and
analyze.

As we have previously noted, models for bubbles in 2+1 (cylinder
systems) with thin shell junction conditions have been
well-studied. For de-Sitter interior, flat exterior the bubble always
recollapses in a finite time.  Even though our models do not satisfy
the thin-wall approximation, the analytic and numerical results for
isolated bubbles suggest that all will eventually become singular.
The observed growth of the Krecschmann scalar is consistent with this
supposition.

When a singularity forms behind an event horizon then one can, in
principle, continue to evolve the solution forward in part of
spacetime.  The hoop conjecture \cite{Thorne:1972ji} suggests that a
horizon forms only when a circular hoop with given size can rotate
freely about the object. In the cylinder system no such hoop
exists. If a singularity forms at some time it will be a naked
singularity and further evolution will be impossible. This situation
is seen in sufficiently prolate axisymmetric collapses
\cite{Shapiro:1991zza}. Singularity formation in the absence of an
apparent horizon is the fundamental limitation to exploring the future
evolution of the 2+1 bubble collisions. By contrast, a collision of
domain walls in 1+1 formed an apparent horizon that covered and
shrouded the singularity \cite{Takamizu:2007ks,Takamizu:2006yd}.

The second issue is more technical. Our method assumes that the deficit angle
is small.  We see no direct evidence that this assumption is violated.
The fact that the constraint equations are poorly satisfied once the
Kretschmann scalar curvature begins to diverge is expected. In the
future we will examine this issue more directly.

Nonetheless, we can begin to make some interesting observations. The
left panel of Fig. \ref{fig-coll3} shows snapshots of the metric for
the collision of two bubbles with different vacua. Initially, $A$
increases at the center of each bubble. At $t \sim 3$ the bubbles
begin to overlap and $A$'s increase slows. Near the center of the
bubble we have $A_{tt}\simeq -0.5A_t^2-0.25\phi_t^2+0.5 e^{2A}V$. The
fact that $A_{tt}<0$ (at the center of bubble) follows because the
magnitude of $A_t$ and $\phi_t$ are large.  The growth of the metric
and the potential in the collapse regions is a multi-step process: we
have seen that the aligned 2+1 collision geometry and the development
of spacetime curvature focuses energy. Part of the field kinetic
energy is transformed into potential energy and enhances $V$. Since
spacetime is not homogeneous the same focusing creates large spatial
gradients in $\phi$ that give rise to large Hubble constants.

The right panel of Fig. \ref{fig-coll3} show the variation of the
field $\phi$ at the same snapshots.  Sections of field with $x>0$
($x<0$) appear to approach the unstable local maxima at $\phi=1.023$
($-0.97$). The field near $x=0$ is constrained to pass through the
global minima of the potential at $\phi=\phi_0$.  At $t \sim 4$ only a
bit of the field return to the local minima on either side of $x=0$
(i.e. $\phi=1.40$ for $x>0$ and $\phi=-1.42$ for $x<0$).  We cannot be
sure if a more violent collision or a different form for $V(\phi)$
might cause more of it to do so. It is apparent that significant parts
of the collision region remain close to the unstable local maxima that
interpolates between the global and local minima.

Fig. \ref{fig-coll5} shows analogous results when bubbles having the
same vacua collide. Now the field near $x=0$ is not constrained to
pass from positive to negative $\phi$. Instead, $\phi$ is close to the
global minimum of the potential at the symmetry point initially. In
the collision the field rises, appears to overshoot the local minima
$\phi=1.40$ and then relaxes back to it. The outer regions interpolate
between the local minima and local maxima of the potential and this
appears to increase in size.

\section{Implications}

We have explored bubbles in cylindrical symmetry with scalar field
stress energy tensors and non-trivial potentials focusing on those
with false vacuum inside and flat (nearly Minkowski) space
outside. The traditional thin shell description omits the transition
region, a part of the spacetime that forms the bridge between
de-Sitter interior and Minkowski exterior.  The transition region
inflates at a lesser rate than the interior of the bubble.  In our
coordinates the thin shell collapses even while the bridge expands.

Nearby expanding de-Sitter bubbles may collide and create spatially
varying vacuum energy density. We evolve examples of colliding systems
until a naked singularity forms. At the end of the simulation
different local observers measure different vacuum energy densities
(effective cosmological constants) in different patches of the
universe. Local potential maxima in field space appear to be
attractors in the sense that the disturbed fields appear to be poised
at or near these maxima. The range in vacuum energy densities is
enhanced when different vacua collide. This dynamical mechanism will
generally introduce inhomogeneous vacuum energy values, the size of
those differences set by the scale between local potential
minima/maxima that lie near each other in field space and that are
accessible via bubble collisions. It is natural to speculate that the
late time appearance of a very small cosmological constant in the
universe might be related to this mechanism.

An important caveat for interpreting these results is that spherical
bubbles may form horizons whereas cylinder bubbles do not. Part of the
interesting dynamics of the cylinder bubble collisions may end up
being shielded from outside observers in spherical systems when
horizons form. This is an issue that cannot easily be addressed within
the context of 2+1 simulations but makes more elaborate 3+1
simulations of great interest.

\acknowledgments
YT would like to thank Masayuki Umemura for discussions on this work. 
YT also wishes to acknowledge financial supports by a Grant-in-Aid through JSPS Fellow for Research Abroad H26-No.27 and by Research Core for the History of the Universe, University of Tsukuba. This material is based upon work supported by the National Science Foundation under Grant No. 1417132.
DC acknowledges NSF's support and the
hospitality of Prof. John Barrow, DAMTP and Clare Hall, University of Cambridge.
We thank DAMTP, the Centre for Theoretical Cosmology, University of Cambridge, where this work was started.


\begin{thebibliography}{99}
\bibitem{Susskind:2003kw} 
  L.~Susskind,
  In {\it Universe or multiverse?} ed by B. Carr, pp247-266, 
hep-th/0302219. 

\bibitem{Langer:1969bc} 
  J.~S.~Langer,
  Annals Phys.\  {\bf 54}, 258 (1969).

\bibitem{Coleman:1977py} 
  S.~R.~Coleman,
  Phys.\ Rev.\ D {\bf 15}, 2929 (1977)
  Erratum: [Phys.\ Rev.\ D {\bf 16}, 1248 (1977)].

\bibitem{Coleman:1980aw} 
  S.~R.~Coleman and F.~De Luccia,
  Phys.\ Rev.\ D {\bf 21}, 3305 (1980). 

\bibitem{Hawking:1982my} 
  S.~W.~Hawking and I.~G.~Moss,
  Nucl.\ Phys.\ B {\bf 224}, 180 (1983).

\bibitem{Guth:1979bh} 
  A.~H.~Guth and S.~H.~H.~Tye,
  Phys.\ Rev.\ Lett.\  {\bf 44}, 631 (1980)
  Erratum: [Phys.\ Rev.\ Lett.\  {\bf 44}, 963 (1980)].
\bibitem{Kazanas:1980tx} 
  D.~Kazanas,
  Astrophys.\ J.\  {\bf 241}, L59 (1980).
\bibitem{Sato:1981ds} 
  K.~Sato,
  Phys.\ Lett.\  {\bf 99B}, 66 (1981).
\bibitem{Guth:1980zm} 
  A.~H.~Guth,
  Phys.\ Rev.\ D {\bf 23}, 347 (1981).
\bibitem{Guth:1982pn} 
  A.~H.~Guth and E.~J.~Weinberg,
  Nucl.\ Phys.\ B {\bf 212}, 321 (1983).

\bibitem{Polchinski:2006gy} 
  J.~Polchinski,
  hep-th/0603249.

\bibitem{Lee:1987qc} 
  K.~M.~Lee and E.~J.~Weinberg,
  Phys.\ Rev.\ D {\bf 36}, 1088 (1987). 

\bibitem{Basu:1991ig} 
  R.~Basu, A.~H.~Guth and A.~Vilenkin,
  Phys.\ Rev.\ D {\bf 44}, 340 (1991).

\bibitem{Hawking:1982ga} 
  S.~W.~Hawking, I.~G.~Moss and J.~M.~Stewart,
  Phys.\ Rev.\ D {\bf 26}, 2681 (1982).

\bibitem{Easther:2009ft} 
  R.~Easther, J.~T.~Giblin, Jr, L.~Hui and E.~A.~Lim,
  Phys.\ Rev.\ D {\bf 80}, 123519 (2009), 
  arXiv:0907.3234.


\bibitem{Giblin:2010bd} 
  J.~T.~Giblin, Jr, L.~Hui, E.~A.~Lim and I.~S.~Yang,
  Phys.\ Rev.\ D {\bf 82}, 045019 (2010), 
  arXiv:1005.3493.


\bibitem{Hwang:2012pj} 
  D.~i.~Hwang, B.~H.~Lee, W.~Lee and D.~h.~Yeom,
  JCAP {\bf 1207}, 003 (2012), 
  arXiv:1201.6109.

\bibitem{Hwang:2014cqa} 
  D.~I.~Hwang, B.~H.~Lee, W.~Lee and D.~H.~Yeom,
  Nucl.\ Phys.\ Proc.\ Suppl.\  {\bf 246-247}, 196 (2014).

\bibitem{Aguirre:2007an} 
  A.~Aguirre, M.~C.~Johnson and A.~Shomer,
  Phys.\ Rev.\ D {\bf 76}, 063509 (2007), arXiv:0704.3473.

\bibitem{Aguirre:2009ug} 
  A.~Aguirre and M.~C.~Johnson,
  Rept.\ Prog.\ Phys.\  {\bf 74}, 074901 (2011), 
  arXiv:0908.4105.

\bibitem{Wainwright:2013lea} 
  C.~L.~Wainwright, M.~C.~Johnson, H.~V.~Peiris, A.~Aguirre, L.~Lehner and S.~L.~Liebling,
  JCAP {\bf 1403}, 030 (2014), 
  arXiv:1312.1357.

\bibitem{Wainwright:2014pta} 
  C.~L.~Wainwright, M.~C.~Johnson, A.~Aguirre and H.~V.~Peiris,
  JCAP {\bf 1410}, no. 10, 024 (2014), 
 arXiv:1407.2950.
\bibitem{Johnson:2015gma} 
  M.~C.~Johnson, C.~L.~Wainwright, A.~Aguirre and H.~V.~Peiris,
  JCAP {\bf 1607}, no. 07, 020 (2016), 
  arXiv:1508.03641.
\bibitem{Kleban:2011pg} 
  M.~Kleban,
  Class.\ Quant.\ Grav.\  {\bf 28}, 204008 (2011), 
  arXiv:1107.2593.

\bibitem{Parry:2012ku} 
  A.~R.~Parry,
  Anal.\ Math.\ Phys.\  {\bf 4}, no. 4, 333 (2014), 
  arXiv:1210.5269.

\bibitem{Thorne65} 
  K.~S.~Thorne,\ 
  Phys.\ Rev.  {\bf 138}, no. 1B, B251 (1965). 


\bibitem{Berezin:1982ur} 
  V.~A.~Berezin, V.~A.~Kuzmin and I.~I.~Tkachev,
  Phys.\ Lett.\  {\bf 120B}, 91 (1983).

\bibitem{Maeda:1985ye} 
  K.~Maeda,
  Gen.\ Rel.\ Grav.\  {\bf 18}, 931 (1986).

\bibitem{Sato:1986xxxx}
  H.~Sato,
  Prog.\ Theor.\ Phys. {\bf 76}, 1250 (1986).

\bibitem{Blau:1986cw} 
  S.~K.~Blau, E.~I.~Guendelman and A.~H.~Guth,
  Phys.\ Rev.\ D {\bf 35}, 1747 (1987).

\bibitem{Berezin:1987bc} 
  V.~A.~Berezin, V.~A.~Kuzmin and I.~I.~Tkachev,
  Phys.\ Rev.\ D {\bf 36}, 2919 (1987).

\bibitem{Aurilia:1989sb} 
  A.~Aurilia, M.~Palmer and E.~Spallucci,
  Phys.\ Rev.\ D {\bf 40}, 2511 (1989).

\bibitem{Suzuki:1991yk} 
  H.~Suzuki, Y.~Fujiwara, T.~Mishima and A.~Hosoya,
  Prog.\ Theor.\ Phys.\  {\bf 86}, 411 (1991).

\bibitem{Sato:1981gv} 
  K.~Sato, H.~Kodama, M.~Sasaki and K.~Maeda, 
  Phys.\ Lett.\ B {\bf 108}, 103 (1982).

\bibitem{Thorne:1972ji} 
  K.~S.~Thorne,
  In *J R Klauder, Magic Without Magic*, San Francisco 1972, 231-258

\bibitem{Shapiro:1991zza} 
  S.~L.~Shapiro and S.~A.~Teukolsky,
  Phys.\ Rev.\ Lett.\  {\bf 66}, 994 (1991).

\bibitem{Takamizu:2006yd} 
Y.~i.~Takamizu and K.~i.~Maeda, 
  Phys.\ Rev.\ D {\bf 73}, 103508 (2006). 

\bibitem{Takamizu:2007ks} 
  Y.~i.~Takamizu, H.~Kudoh and K.~i.~Maeda,
  Phys.\ Rev.\ D {\bf 75}, 061304 (2007), 
  gr-qc/0702138.

\end{thebibliography}
\end{document}